\documentclass[twocolumn,showpacs,graphicx,floatfix,eqsecnum,prb]{revtex4-1}
\usepackage{amssymb}
\newcommand{\be}{\begin{equation}}
\newcommand{\ee}{\end{equation}}
\newcommand{\bea}{\begin{eqnarray}}
\newcommand{\eea}{\end{eqnarray}}
\newcommand{\bwt}{\begin{widetext}}
\newcommand{\ewt}{\end{widetext}}
\newcommand{\bk}{{\bf k}}
\newcommand{\bq}{{\bf q}}
\newcommand{\bp}{{\bf p}}
\newcommand{\bkp}{{\bf k}'}

\newcommand{\bv}{{\bf v}}
\newcommand{\bu}{{\bf u}}
\newcommand{\ek}{\varepsilon_{\mathbf k}}
\newcommand{\ekp}{\varepsilon_{{\mathbf k}'}}

\newcommand{\I}{\mathrm{Im}}
\newcommand{\R}{\mathrm{Re}}
\newcommand{\bsu}{\begin{subequations}}
\newcommand{\esu}{\end{subequations}}
\newcommand{\ho}{$\mathrm{URu}_2\mathrm{Si}_2$}
\newcommand{\ce}{$\mathrm{Ce}_{0.095}\mathrm{Ca}_{0.05}\mathrm{TiO}_{3.04}$}
\newcommand{\nd}{$\mathrm{Nd}_{0.905}\mathrm{TiO}_3$}

\usepackage{amsmath}
\usepackage{epsf}
\usepackage[latin9]{inputenc}
\usepackage{amsmath}
\usepackage{amssymb}
\usepackage{graphicx}
\usepackage{esint}
\usepackage{epsf}

\begin{document}

\title{First-Matsubara-frequency
 rule
  in a Fermi liquid. \\ Part II: Optical conductivity  and comparison to experiment}

\author{ Dmitrii L. Maslov$^1$ and Andrey V. Chubukov$^2$}

\affiliation{$^1$ Department of Physics, University of Florida, P.O. Box 118440, Gainesville, FL 32611-8440\\
$^2$ Department of Physics, University of Wisconsin-Madison, 1150 Univ. Ave., Madison, WI 53706-1390}

\begin{abstract}
Motivated by recent optical measurements on a number of strongly correlated electron systems,
 we revisit
  the dependence
  of the conductivity of a Fermi liquid,
 $\sigma(\Omega,T)$,
 on the
  frequency $\Omega$ and temperature $T$.
Using the Kubo formalism
 and taking full account of vertex corrections, we show that the
  Fermi liquid
  form $\R\sigma^{-1}(\Omega,T)\propto \Omega^2+4\pi^2T^2$
holds under very general conditions, namely
in any dimensionality above one,
 for a Fermi surface of an arbitrary shape (but away from nesting and van Hove singularities),
and to any order in the
electron-electron interaction.
We also show that
 the scaling form of $\R\sigma^{-1}(\Omega,T)$
is
 determined by
  the
 analytic properties of the conductivity along the Matsubara axis. If a system contains not only itinerant electrons
but also localized degrees of freedom
 which scatter electrons elastically, e.g., magnetic moments or resonant levels, the scaling form changes to $\R\sigma^{-1}(\Omega,T)\propto \Omega^2+b\pi^2T^2$,
  with $1\leq b< \infty$. For purely elastic scattering,
 $b =1$.
 Our analysis implies that
 the value of $b\approx 1$, reported for \ho\; and some
 rare-earth based
  doped Mott insulators,
  indicates that
  the optical conductivity
  in these materials
   is controlled by an elastic scattering mechanism,
   whereas
   the values of
        $b\approx 2.3$ and $b\approx 5.6$,
     reported for underdoped cuprates and organics,
      correspondingly, imply
      that
  both elastic and inelastic
   mechanisms contribute to the optical conductivity.
\end{abstract}

\maketitle

\section{introduction}

Optical response of strongly correlated materials is an invaluable tool for studying
the
 dynamics of charge carriers.~\cite{basov:11}
On par with the angular-resolved photoemission spectroscopy
 which
    gives
    information about the single-particle self-energy, optical experiments provide information about two other important quantities: the dynamical effective mass and the scattering rate
    of conduction electrons.

 In the preceding paper\cite{CMI} (hereafter referred to as I), we discussed  constraints imposed  on the functional form of the
 retarded single-particle self-energy $\Sigma^R
 (\omega, T)$
  by the \lq\lq first-Matsubara-frequency rule\rq\rq\/.
 This rule stipulates
 that,
 under certain conditions,
    a function obtained by analytic continuation of $\I\Sigma^R(\omega,T)$
 to the Matsubara
 axis must vanish at
 the first fermionic Matsubara frequency $\omega\to \pm i\pi T$.
 The
 familiar  scaling form of $\I\Sigma^R(\omega,T)$ in a generic Fermi liquid (FL) in $D >2$
 \be
 \I\Sigma^R(\omega,T)=C\left(\omega^2+\pi^2T^2\right),
 \label{cfl}
 \ee
 (with a coefficient of exactly $\pi^2$ in front of the $T^2$ term),
 obviously satisfies this rule.

 In the present paper, we discuss a similar
 constraint--\lq\lq the first bosonic Matsubara-frequency rule\rq\rq\/-- imposed on the scaling form of the
   optical conductivity $\sigma(\Omega, T)$.

Within the semiclassical Boltzmann equation, the $T^2$-scaling of the {\em dc} resistivity 
due to Umklapp electron-electron scattering
was obtained by Landau and Pomeranchuk,\cite{landau:36} and due normal scattering in a two-band metal by Baber.\cite{baber}  Later on, Eliashberg~\cite{eliash:62} re-derived this result from the Kubo formula, and showed that it remains valid to all orders in the electron-electron interaction.
 The $\Omega/T$ scaling of the  \lq\lq optical resistivity\rq\rq\/ of a
 3D
  FL
  was first discussed by Gurzhi,~\cite{gurzhi:58}  who used a quantum Boltzmann equation to show that
 \be
\R \left[\rho(\Omega,T)\right]\equiv \R\left[\sigma^{-1}(\Omega, T)\right] =A'\left[\Omega^2+4\pi^2T^2\right].
\label{gurzhi}
\ee
The $\Omega$ and $T$ dependences of $\R~\rho(\Omega, T)$ are similar to those of
 the leading term in
$\mathrm{Im} \Sigma^R(\omega,
T)$ [cf.~Eq.~(\ref{cfl})], but the ratio of the $T^2$
and $\Omega^2$ terms is now $4\pi^2$ instead of $\pi^2$.

This difference is not accidental.
Indeed, $\I\Sigma^R(\omega,T)$ measures the decay rate of single-particle excitations, which are fermions;
hence the thermal part of
$\I\Sigma^R(\omega,T)$ contains the square  of first fermionic Matsubara frequency ($=\pi T$) rather than $T$ itself.
On the other hand, $\R
\rho(\Omega, T)$ measures the decay rate of current fluctuations, which are bosons;
 hence the thermal part of $\R\sigma(\Omega,T)$ contains the square of the  first (non-zero) bosonic Matsubara frequency ($=2\pi T$). Also not coincidentally, Eq.~(\ref{gurzhi}) is of the same form as the sound absorption rate in FLs.~\cite{physkin}

To the best of our knowledge, the scaling form predicted by Eq.~(\ref{gurzhi}) has never been verified experimentally in conventional
metals. On the other hand, the $\Omega/T$ scaling of $\R\rho(\Omega,T)$
has been studied intensively in strongly correlated materials, e.g., in heavy-fermion metals and doped Mott insulators.
The result of these studies is quite surprising: whenever it was possible to fit the $\Omega$ and $T$
 dependencies
  of   $\R\rho(\Omega,T)$ by quadratic functions, the coefficient $b\equiv \pi^2T^2/\Omega^2$  was
found to be quite different from $4$.
This issue was highlighted by recent study~\cite{timusk11} of  $\R\rho (\Omega, T)$
in the \lq\lq hidden-order\rq\rq\/ (HO) heavy-fermion compound URu$_2$Si$_2$, where $b$ was found to be close to $1$ above the $17.5$ K  transition to the HO state. In fact, the
 value of
$b\approx 1$
 was found in a number of other materials, including
two rare-earth based
 doped Mott insulators, \nd~(Ref.~\onlinecite{tio3}) and \ce~(Ref.~\onlinecite{ndtio3}).
 Another
  recent study~\cite{mirzaei}
  reports
    $b\approx 2.3$ in the underdoped cuprate HgBa$_2$CuO$_{4+\delta}$.
Whereas the observed value of $b$ is less than $4$ in most of the cases, there is one exception:
$b\approx 5.6$ was reported for a quasi-two-dimensional (2D)
 organic material of the BEDT-TTF family.\cite{dressel}
 A deviation of $b$ from $4$ can also be inferred from the optical data on UPt$_3$,\cite{sulewski88} Sr$_2$RuO$_4$,\cite{sr2ruo4}and Cr;\cite{basov_cr}
see Ref.~\onlinecite{timusk11}  for more details.

 Motivated by these findings,  we revisit the $\Omega/T$ scaling of the optical conductivity of a FL in this paper.
 We
  extend the Eliashberg's analysis of the Kubo formula for the conductivity to finite $\Omega$
  and obtain an expression for $\sigma(\Omega,T)$ to all orders in the electron-electron interaction.
   To discuss the results,
  it is convenient to identify two distinct frequency regimes, the \lq\lq high-frequency\rq\rq\/ and \lq\lq low-frequency\rq\rq\/ ones,
  and also two types of
   FLs, the \lq\lq conventional\rq\rq\/ and \lq\lq non-conventional\rq\rq\/ ones.

  As far as the frequency regimes are concerned,
    $\Omega$ is larger than $\I \Sigma^R(\Omega, T)$ in the high-frequency regime,
    while  $\Omega < \mathrm{Im} \Sigma^R (\Omega, T)$ in the  low-frequency  one.
     (The low-frequency regime also includes the {\em dc} limit of $\Omega=0$.)
    In the context of the standard Drude formula, these regimes are also referred to as \lq\lq non-dissipative\rq\rq\/ or \lq\lq reactive\rq\rq\/ and \lq\lq dissipative\rq\rq\/, correspondingly.

    Turning to two types of FLs,
we define a \lq\lq conventional FL\rq\rq\/ as such in which the leading $\omega$ and $T$ dependencies of $\I\Sigma^R
 (\omega, T)$ are given by $\omega^2 + \pi^2T^2$, as in  Eq.~(\ref{cfl}), while the higher-order terms may be non-analytic. In a \lq\lq nonconventional FL\rq\rq\/,
already the leading term in $\I\Sigma^R(\omega, T)$ is  a non-analytic function of $\omega$ and $T$.
For a wide class of interactions that
 remain finite at $q=0$, the demarcation line between the two types of FLs is determined by the dimensionality: the case of $D>2$ corresponds to conventional FLs, while the case of
 $1<D<2$
 corresponds to non-conventional FLs. In the latter case,
 $\I\Sigma^R(\omega,0)\propto |\omega|^D$
and $\I\Sigma^R(0,T)\propto T^D$.
In the marginal case of $D=2$,
 $\I\Sigma^R(\omega,0)\propto \omega^2\ln|\omega|$
and
$\I\Sigma^R(0,T)\propto T^2\ln T$.

 In the high-frequency regime,
   current-carrying quasiparticles can be considered as nearly free, so that
   the residual interaction among quasiparticles, which gives rise to their finite lifetime, can be treated as a perturbation.
 We show that in this regime $\R\rho(\Omega,T)$ is
 given by Eq.~(\ref{gurzhi}) for both conventional and non-conventional FLs,
 as well as
 for the marginal case of  $D=2$,
  despite
  qualitative differences in the self-energies
  in these cases.
   Our analysis keeps full track of the vertex corrections to the conductivity
  and thus takes both normal and Umklapp scattering processes into account.
    We argue that the
     $4\pi^2$
    coefficient
    of the $T^2$ term in this formula is
     a consequence of the  \lq\lq bosonic first-Matsubara-frequency rule\rq\rq\/,
     which stipulates  that a function
     obtained by  analytic continuation of
     $\R\rho(\Omega,T)$
     to  the first (non-zero) bosonic Matsubara frequency, $\Omega\to \pm 2i\pi T$,
     does not have a $T^2$ term.
      Equation (\ref{gurzhi}) obviously obeys this rule.

  In the low-frequency regime, $\R\rho(\Omega, T)$ differs from Eq.~(\ref{gurzhi})
  because the interaction among quasiparticles
   can
   no longer
    be treated as a perturbation,
        and this affects the $T^2$ and $\Omega^2$ terms in
  $\R\rho(\Omega, T)$ in different ways.
   We analyzed the change in the functional form of $\rho (\Omega, T)$ between the high- and low-frequency regimes in the
      \lq\lq zero-bubble approximation\rq\rq\/ \cite{georges:96}
 and found that the change is numerically quite small, i.e.,
 the formula
 \be
 \R\rho(\Omega,T)=A'\left(\Omega^2 + b \pi^2 T^2\right)
 \label{pheno}
 \ee
  with $b \approx 4$ remains quite accurate
   down to the lowest $\Omega$, although the exact form of
   $\R \rho(\Omega, T)$ in the entire range of $\Omega$
   is different from that in Eq.~(\ref{gurzhi}).
    We also analyzed $\R \rho(\Omega, T)$ in the \lq\lq incoherent regime\rq\rq\/, where
    all energy scales are of the same order, i.e.,
    $\Omega \sim T
    \sim\R\Sigma^R(\Omega,T)\sim
    \I\Sigma^R(\Omega, T)$, and again found a good fit by the $\Omega^2 + b \pi^2 T^2$ form with $b \approx 4$.

 Equation~(\ref{pheno}) is to be taken with some caution, because
the zero-bubble approximation neglects the corrections to the current vertex in the polarization bubble.
 Physically,
 vertex corrections differentiate between normal and Umklapp scattering processes.
 In the high-frequency regime, both normal and Umklapp processes contribute to the
 resistivity
provided that the Fermi surface (FS) is sufficiently anisotropic~\cite{gurzhi:58,rosch05,maslov1} (a precise definition
 of \lq\lq sufficiently anisotropic\rq\rq\/ is given in Sec.~\ref{sec:5}).
As a result, 
 Eq.~(\ref{gurzhi})
  remains valid 
    when the vertex corrections 
  are included. 
The only change is that the prefactor $A'$ now contains a sum of normal and Umklapp scattering amplitudes.
In the low-frequency regime and, in particular, at $\Omega =0$, the
   resistivity of an impurity-free system is non-zero only in the presence of Umklapp scattering,
   although normal processes also contribute once Umklapp processes are allowed.~\cite{maebashi97_98}
   As  a result, the prefactor $A$ in the {\em dc} resistivity
   \be
\rho(0,T)=4\pi^2 AT^2
\label{dca}
\ee
contains some function of the normal and Umklapp scattering amplitudes rather than just their sum,
and is therefore different from $A'$ in the high-frequency limit.
 What remains to be seen is how  the functional form of $\R\rho(\Omega,T)$ evolves between the {\em dc} and high-frequency limits beyond the zero-bubble approximation.

  We then
   discuss
   the experiment,
  focusing mostly
    on recent optical measurements on URu$_2$Si$_2$.~\cite{timusk11}
   Given that the observed values of the coefficient $b$
   are
   substantially
   different
    from
   the
   FL value
   $b=4$,  we argue that the existing optical data cannot be explained
    only by
    the electron-electron interaction.
  Following an analogy with the Kondo effect,~\cite{kondo} we propose a phenomenological model which, in addition to  electron-electron scattering, contains also {\em elastic} scattering by some localized decrees of freedom, e.g., magnetic moments or resonant levels.
 In this model, the
 self-energy is a sum
  of two parts: the elastic one, described by an $\omega^2$ term , and the inelastic one, described by the standard FL term, $\omega^2+\pi^2 T^2$, i.e.,
 \be
 \I\Sigma^R(\omega,T)=C\left[a\omega^2+\left(\omega^2+\pi^2T^2\right)\right],
 \label{pheno_se}
 \ee
where the relative weight of the elastic and inelastic contributions, $a$,  is an adjustable parameter of the model.   Within the zero-bubble approximation, the coefficient $b$ in Eq.~(\ref{pheno}) is related to $a$ via
   \be
   b=\frac{a+4}{a+1}.
   \ee
   For practical purposes, the model is
   meaningful
    only for $-1<a<\infty$; consequently, $1\leq b <\infty$.      The  FL value of $b=4$ is reproduced for $a=0$.  The opposite limit of $a=\infty$ (and thus $b=1$) corresponds to a purely elastic scattering mechanism.
   The range $1<b<4$ corresponds to a mixture of elastic and inelastic mechanisms with $a>0$, whereas $b>4$ corresponds to an
  elastic contribution with $-1<a<0$.

    In this classification scheme,
    the value of $b\approx 1$, reported in Refs.~\onlinecite{timusk11,tio3,ndtio3}, indicates a purely elastic scattering mechanism,
    whereas $b\approx 2.3$ (and thus $a\approx 1.3$) and $b\approx 5.6$ (and thus $a\approx -0.35$),  reported in Refs.~\onlinecite{mirzaei} and \onlinecite{dressel}, correspondingly, point at a mixture of elastic and inelastic mechanisms with opposite signs of the elastic contribution.

   We discuss one possible mechanism that leads to $b\approx 1$, i.e.,
   scattering at resonant levels, and show that this mechanism explains the data on \ho  ~reasonably well.
   We refrain from identifying the microscopic origin of the resonant levels (except for noting that {\em extrinsic} resonant impurities can hardly be the culprits) but merely surmise that {\em intrinsic} deep electron states
   can play a role of incoherent resonant scatterers at relatively high
    energies,
    where
    a
    coherent Bloch state is not formed yet.

    Whereas the resonant-level model explains  the optical data in the \lq\lq $b=1$\rq\rq\/ materials,
    the $T$ dependence of the {\em dc} resistivity can be explained only by invoking a sufficiently strong electron-electron interaction which, when combined the resonant elastic scattering,
    does not significantly affect the optical scattering rate.
  We show that
   {\em dc} and optical measurements probe different scattering mechanisms:
   while a {\em dc} measurement is sensitive to both elastic and inelastic mechanisms,
   an optical measurement probes primarily the elastic channel,

    The rest of the paper is organized as follows. In Sec.~\ref{sec:kubo} we discuss the optical conductivity
    of
     a FL in the high-frequency regime within the Kubo formalism.
     In Sec.~\ref{sec:4},   we extend the analysis to both
     the
      low-frequency and incoherent regimes within the
        zero-bubble approximation for the current-current correlator.
      In Sec.~\ref{sec:5}, we discuss the interplay between normal and Umklapp contributions to the resistivity in different frequency regimes.
      Section ~\ref{sec:6} addresses comparison to the experiment. In Sec.~\ref{sec:status},
      we discuss
      the status of the experiment and conclude that it cannot be explained within the model which includes only the electron-electron interaction. In Sec.~\ref{sec:elastic}, we introduce a phenomenological model which combines elastic and inelastic scattering mechanisms, and classify
      the
      observed values of the coefficient $b$ within this model.  In Sec.~\ref{sec:resimp}, we apply the  resonant-scattering model to the data on URu$_2$Si$_2$.  Section~\ref{sec:7} presents our conclusions.

\section{Optical conductivity\\ of a Fermi liquid}
\label{sec:kubo}
\subsection{Kubo formula: rigorous treatment}
\subsubsection{Preliminaries}
As in I, we consider an electron system on a lattice.
 We assume that the FS does not have nested parts and is away from van Hove singularities but is otherwise arbitrary. Near the FS, the bare electron dispersion, $\ek^0$, (measured from the Fermi energy) is approximated by $\ek^0=\bv_{\bk_F}^0\cdot (\bk-\bk_F)$,  where $\bk_F$ is a vector in the direction of $\bk$ and residing on the FS.
Following the conventional FL methodology, we divide electron states into \lq\lq low-energy\rq\rq\/ (near the Fermi energy) and \lq\lq high-energy\rq\rq\/ ones. Effects of the interaction via high-energy states are parameterized by the self-energy ${\tilde \Sigma}^R_{\bk}(\omega)$.
 An expansion of ${\tilde \Sigma}^R_{\bk}(\omega)$ near the FS
\be
{\tilde\Sigma}_{\bk}^R(\omega)=\omega\left(\frac{1}{Z_{\bk_F}}-1\right)+
\left({\bf u}_{\bk_F}
-\bv_{\bk_F}^0\right)\cdot(\bk-\bk_F)
\ee
defines the
 quasiparticle
 renormalization factor
\be
Z_{\bk_F}=\left(1+\frac{\partial{\tilde\Sigma}^R_{\bk_F}}{\partial\omega}{\Big\vert}_{\omega=0}\right)^{-1}\ee
 and renormalized dispersion  $\ek=\bv_{\bk_F}\cdot(\bk-\bk_F)$  of the low-energy states,
 where
  \be
 \bv_{\bk_F}=Z_{\bk_F}{\bf u}_{\bk_F}=Z_{\bk_F}\left(\bv^0_{\bk_F}
 -
 \boldsymbol\nabla_{\bk}{\tilde\Sigma}_{\bk}(0)\left\vert_{\bk=\bk_F}\right.\right).
 \label{uf}
 \ee
 (As in I, we define the
   single-particle
   self-energy
 as $G^{-1}_{\bf k} (\omega, T) = \omega + \Sigma_{\bf k} (\omega, T) - \epsilon_{\bk}$.) 
  The renormalization factor and both velocities ($\bv_{\bk_F}$ and $\bu_{\bk_F}$) are defined at point $\bk_F$ of the FS
 and, in general, vary over the FS.
The (Matsubara) Green's function describing the low-energy electron states is given
by
\bea
G^0_{\bk}(\omega_m)
=
\frac{1}{i\omega_m/Z_{\bk_F}-{\bf u}_{\bk_F}\cdot(\bk-\bk_F)}
 =\frac{Z_{\bk_F}}{i\omega_m-\ek},\notag\\
 \label{bare}
\eea
where $\omega_m=\pi T(2m+1)$.

The combination of properties formulated above defines the \lq\lq bare\rq\rq\/ low-energy theory described by
the
  action
\be
{\cal S}=T\sum_{\omega_m}\int_{\bk} {\bar\psi}_{\omega_m,\bk}\left[i\omega_m/Z_{\bk_F}-{\bf u}_{\bk_F}\cdot(\bk-\bk_F)\right]\psi_{\omega_m,\bk},
\label{action}
\ee
where $\int_{\bk}$ is a shorthand notation for $\int d^Dk/(2\pi)^D$.
The residual interaction between low-energy quasiparticles is described by an instantaneous potential $U_{\bf q}$, which is already dressed up by high-energy states and assumed to be non-singular for
any $\bq$ that connects two points on the FS, including $\bq=0$. Dynamic screening of the interaction by low-energy states, which gives rise to finite lifetime of quasiparticles and hence finite conductivity,
is treated explicitly. To avoid double-counting, we
 assume that mass renormalization is already absorbed into the parameters of the bare theory.

In the presence of an external electromagnetic field described by vector potential ${\cal A}$, the momentum $\bk$ in the  bare action, Eq.~(\ref{action}), is replaced by $\bk-e{\cal A}/c$.
 The corresponding current vertex contains the
\lq\lq charge velocity\rq\rq\/ ${\bf u}_{\bk_F}$ [Eq.~(\ref{uf})].
Note that ${\bf u}_{\bk_F}$  is renormalized only by the $\bk$ dependent part of ${\tilde\Sigma}^R_{\bk}(\omega)$,~\cite{eliash:62,michaeli:09}
in contrast to the full Fermi velocity $\bv_{\bk_F}$, which is renormalized both by
$\boldsymbol\nabla_{\bk}{\tilde\Sigma}^R$ and
$
\partial{\tilde\Sigma}
^R_{\bk_F}/
\partial\omega
$.

\begin{figure}[t]
\includegraphics[width=0.5\textwidth]{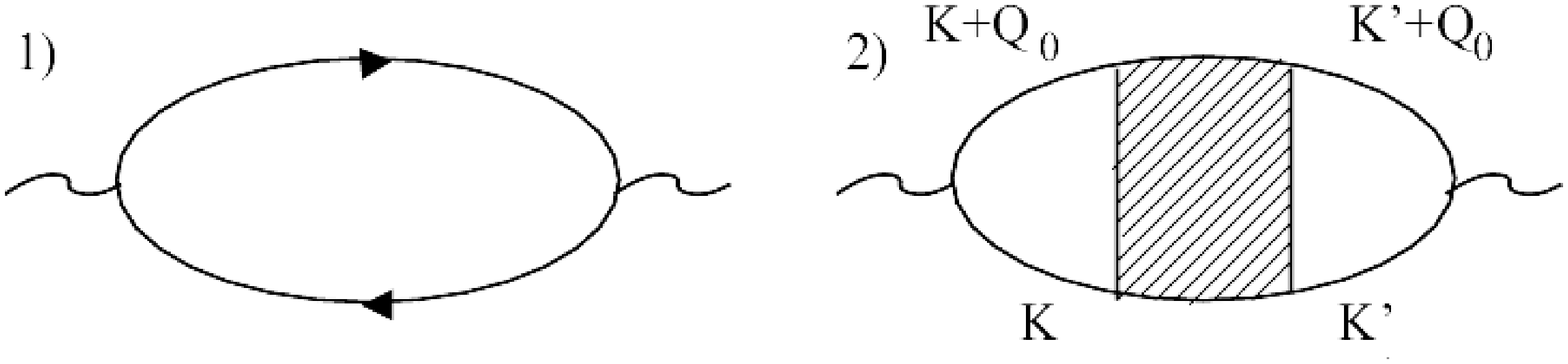}
\caption{Diagrams for
 the current-current correlation function: $\mathcal{K}_1$ (diagram 1)  and $\mathcal{K}_2$ (diagram 2).
The \lq\lq four-momenta\rq\rq\/  in diagram 2) are $K=(\omega_m,\bk)$, $K'=(\omega_{m'},\bkp)$,  and $Q_0=(\Omega_n,0) $. The shaded box is the vertex $\Gamma_{\bk\bkp}(\omega_m,\omega_{m'},\Omega_n)$.
}
\label{fig:kubo}
\end{figure}

To simplify notations, we assume that a metal has cubic symmetry, in which case the conductivity tensor reduces to
$\sigma_{ij}=\delta_{ij}\sigma$. The diagonal component of the conductivity is given by the Kubo formula \be
 \sigma(\Omega,T) = \frac{
 e^2}{
 i\Omega} \left[{\cal K}^{R}(\Omega,T)+{\cal K}^{\mathrm{dia}}(T) \right],
\label{15}
\ee
where ${\cal K}^{R}(\Omega,T)={\cal K}^{R}_1(\Omega,T)+{\cal K}^{R}_2(\Omega,T)$ is the retarded current-current correlation function, represented by a sum of two diagrams in Fig.~\ref{fig:kubo}, and $(e^2/i\Omega) {\cal K}^{\mathrm{dia}}(T)$ is the diamagnetic part of the conductivity,
which cancels the $\Omega =0$ term in  ${\cal K}^{R}(\Omega,T)$  [the sum ${\cal K}^{R}(0,T)+{\cal K}^{\mathrm{dia}}(T)$ must vanish for a normal metal by gauge invariance].
In what follows, we assume that the $\Omega=0$ piece is already subtracted from ${\cal K}^{R}(\Omega,T)$
and do not
specify
 an explicit form
 of ${\cal K}^{\mathrm{dia}}$.

 On the Matsubara axis, the diagrams in Fig.~\ref{fig:kubo} are given by
 \bea
{\cal K}_1(\Omega_n,T)=
-\frac{2}{D} T\sum_{\omega_m}
\int_\bk u^2_{\bk_F} G_{\bk}(\omega_m) G _{\bk}(\omega_m+\Omega_n)\notag\\
\label{k1}\eea
and
\bwt
\bea
 {\cal K}_2(\Omega_n,T)=-
 \frac{2}{D} T^2\sum_{\omega_m,\omega_{m'}}
\int_\bk\int_{\bkp} \bu_{\bk_F}\cdot\bu_{{\bk}'_F}G_{\bk}(\omega_m) G _{\bk}(\omega_m+\Omega_n)\Gamma_{\bk\bkp}(\omega_m,\omega_{m'},\Omega_n) G_{\bkp}(\omega_{m'})
 G_{\bkp}(\omega_{m'}+\Omega_n),
 \label{16}
 \eea
 \ewt
where the Green's functions and the vertex part $\Gamma_{\bk\bkp}(\omega_m,\omega_{m'},\Omega_n) $ contain the effects of residual interaction between low-energy quasiparticles.

The calculation of diagram 1 in Fig.~\ref{fig:kubo} is
 fairly
 straightforward. Replacing the Matsubara sum by a contour integral and converting the momentum integral into integrals over $d\ek$ and over the FS element $dA_{{\bf k}_F}$, we obtain for the imaginary part of ${\cal K}^R_1$
\bea
&&\I{\cal K}_1^R(\Omega,T)=\frac{2}{\pi D(2\pi)^D}\oint dA_{\bk_F}\frac{u^2_{\bk_F}}{v_{\bk_F}} \int d\omega\int d\ek\notag\\
&&\times\left[n_F(\omega)-n_F(\omega+\Omega\right]\I G^R_{\bk}(\omega)\I G^R_{\bk}(\omega+\Omega),
\label{k1}\eea
where $n_F(\epsilon)$ is the Fermi function, $G_{\bk}^{R,A}(\omega) =Z_{\bk_F}/[\omega-\ek\pm iZ_{\bk_F}\I\Sigma^{R}_{\bk}(\omega)]$, and $\I\Sigma^{R}_{\bk}(\omega)$ accounts
for the residual interaction.

\subsubsection{Canonical Fermi liquids}
\label{sec:cfl}

In this Section, we restrict the analysis to conventional FLs.
(We  will show later, in Sec.~\ref{sec:ncfl}, that the result for the conductivity applies to non-conventional FLs as well).
For a conventional FL on the lattice, $\I\Sigma^R_{\bk_F}(\omega,T)$ is still given by Eq.~(\ref{cfl}) with the only
proviso
   that the prefactor now varies along the FS: $C\to C_{\bk_F}$. The dependence of   $\I\Sigma^R_{\bk_k}(\omega,T)$ on $\ek$ is weak and can be neglected. The integral over $\ek$ is then
    solved readily:
\bea
&&\int d\ek \I G^R_{\bk}(\omega)\I G^R_{\bk}(\omega+\Omega)\label{ek}\\
&&=\pi Z_{\bk_F}\I \left[\Omega/Z_{\bk_F}+i\I\Sigma^R_{\bk_F}(\omega,T)+i\I\Sigma^R_{\bk_F}(\omega+\Omega,T)\right]^{-1}.\notag
\eea
The high-frequency regime is defined by the condition
\be
\Omega\gg Z_{\bk_F}\I\Sigma_{\bk_F}(\Omega,T).
\label{hf}
\ee
For a conventional FL,  this condition implies that $\Omega \gg C_{\bk_F} \max\{\Omega^2,T^2\}$ for all points on the FS. The relation between $\Omega$ and $T$ is arbitrary
 but we do
 assume that $\Omega, T \ll E_F$.
In this regime,
Eq.~(\ref{ek}) is expanded in the imaginary parts of the self-energies and their sum is averaged with the difference of the Fermi functions. For a conventional FL, the last step amounts to
\bea
&&\int^{\infty}_{-\infty}d\omega \left[n_F(\omega)-n_F(\omega+\Omega)\right]\left[\I\Sigma^R_{\bk_F}(\omega)+\I\Sigma^R_{\bk_F}(\omega+\Omega)\right]\notag\\
&&=C_{\bk_F}\!\!\int^{\infty}_{-\infty}\!\!d\omega\left[n_F(\omega)\!-\!n_F(\omega+\Omega)\right]\left[\omega^2+(\omega+\Omega)^2+2(\pi T)^2\right]\notag\\
&&=\frac{2}{3}C_{\bk_F}
\Omega
\left(\Omega^2+4\pi^2 T^2\right).\label{4pi2}
\eea
It is at this step when the difference between the coefficients of the $T^2$ parts in $\I\Sigma^R_{\bk_F}$ and $\sigma$ occurs.
Using (\ref{4pi2}), we obtain
\bea
\R\sigma_1(\Omega,T)=\frac{e^2}{\Omega}\I{\cal K}_1^R(\Omega,T)=B_1\frac{\Omega^2+4\pi^2T^2}{\Omega^2}\notag\\
\label{sigma1}
\eea
with $B_1=(4e^2/3 D(2\pi)^D)\oint dA_{\bk_F}\left(u^2_{\bk_F}/v_{\bk_F}\right)Z_{\bk_F}^3C_{\bk_F}$.
In the high-frequency regime, $\R\sigma\ll \I\sigma=\omega_p^2/4\pi\Omega$, where $\omega_p$ is the effective plasma frequency.
  Expanding $\rho(\Omega,T)=1/\sigma(\Omega,T)$ in $\R\sigma/\I\sigma$, we obtain Eq.~(\ref{gurzhi}) with prefactor $A'=(4\pi)^2B_1/\omega^4_p$.

To analyze the contribution of the vertex corrections represented by diagram 2 in Fig.~\ref{fig:kubo}, we perform analytic continuation of ${\cal K}_2(\Omega_n,T)$, following the procedure developed by Eliashberg.~\cite{eliash:62}  The resulting expression is quite involved but
 to find the real part the conductivity
  we need only that part of ${\cal K}^R_2(\Omega,T)$ which contains the product of the retarded and advanced Green's functions located
  on the same side
   relative to
  the vertex. Only such products
 will survive upon integrating over $\ek$ and $\ekp$. In general, ${\cal K}
^R_2(\Omega,T)$ contains vertices
  which
  are obtained by analytically
  continuation
   the Matsubara vertex $\Gamma_{\bk\bkp}(\omega_m,\omega_{m'},\Omega_n)$ via the following relations: $i\omega_m=\omega+i\I\omega$, $i\omega_{m'}=\omega'+i\I\omega'$, and $i\Omega_n=\Omega+i\I\Omega$, where all the imaginary parts are infinitesimally small.  Analytic properties of continued vertices are determined by relations between the imaginary parts of the three frequencies. The part of ${\cal K}
^R_2(\Omega,T)$ that we are interested in contains vertices $\Gamma^{\mathrm{II}-\mathrm{IV}}_{\bk\bkp}(\omega,\omega',\Omega)$, where Roman numerals indicate regions in the
 $(\I\omega$, $\I\omega')$ plane, as
  shown in Fig.~\ref{fig:regions} (for definiteness, we set $\I\Omega>0$).
 Explicitly,
\bwt
 \bea
 {\cal K}^R_2(\Omega,T)=\frac{1}{4\pi^2D}\int_{\bk}\int_{\bkp}\bu_{\bk_F}\cdot\bu_{\bkp_F}\int d\omega\int d\omega'\left[
 n_F(\omega)-n_F(\omega+\Omega)\right]G_\bk^R(\omega+\Omega)G_\bk^A(\omega)\Gamma_{\bk\bkp}(\omega,\omega',\Omega)G_{\bkp}^R(\omega'+\Omega)G^A_{\bkp}(\omega'),
 \notag\\\label{vert2}
 \eea
where
 \bea
 \Gamma_{\bk\bkp}=\coth\frac{\omega'-\omega}{2T}\left(\Gamma^{\mathrm{II}}_{\bk\bkp}-\Gamma^{\mathrm{III}}_{\bk\bkp}\right)+\coth\frac{\omega+\omega'+\Omega}{2T}\left(\Gamma_{\bk\bkp}^{\mathrm{III}}-\Gamma_{\bk\bkp}^{\mathrm{IV}}\right)-\tanh\frac{\omega'}{2T}\Gamma_{\bk\bkp}^{\mathrm{II}}+\tanh\frac{\omega'+\Omega}{2T}\Gamma_{\bk\bkp}^{\mathrm{IV}}.
 \label{vert1}
 \eea
 \ewt
 [For brevity, we do not spell out the arguments $\omega,\omega',\Omega$ which are the same in all
  vertices in Eq.~(\ref{vert1}).]
   \begin{figure}[t]
\includegraphics[width=0.5\textwidth]{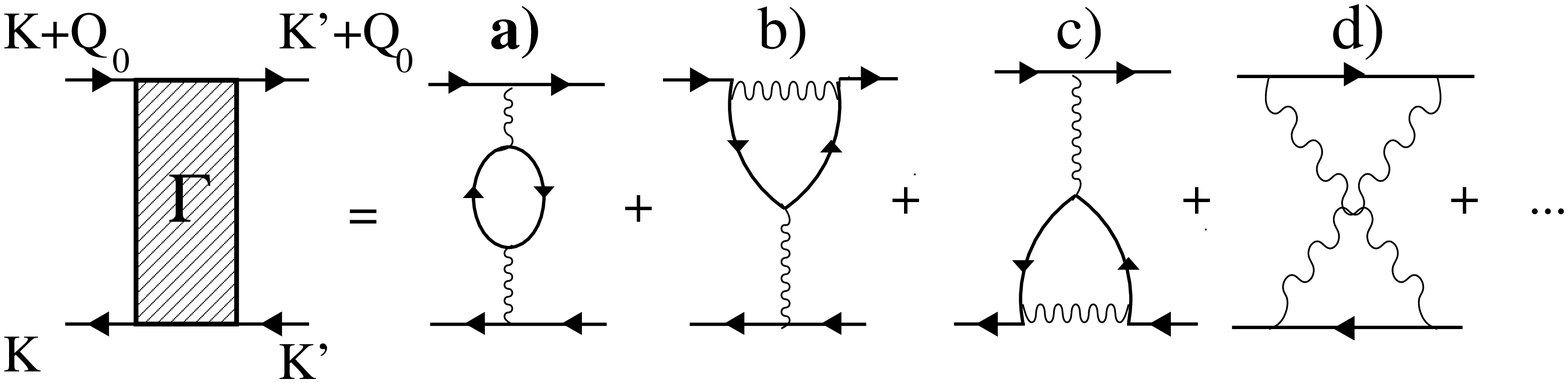}
\caption{
Lowest order diagrams for the vertex in diagram 2) of Fig.~\ref{fig:kubo}.}
\label{fig:vertex}
\end{figure}

Equations (\ref{vert1}) and (\ref{vert2}) allow one to extract the $\Omega$ and $T$ dependences for any vertex diagram.
For example, vertex diagram $a$ in Fig.~\ref{fig:vertex} reads
\bea
&&\Gamma^{\{a\}}_{\bk\bkp}(\omega_m,\omega_{m'},\Omega_n)=U^2_{\bk-\bkp}\Pi_{\bk-\bkp}(\omega_m-\omega_{m'}),
\eea
where $\Pi_{\bq}(\epsilon)$ is the polarization bubble. Continuing this expression to real frequencies, we obtain
$\Gamma_{\bk\bkp}^{a, \mathrm{II}}=U_{\bk-\bkp}^2\Pi_{\bk-\bkp}^R(\omega-\omega')$ in region II, where $\I(\omega-\omega')>0$, and
 $\Gamma_{\bk\bkp}^{a,\mathrm{III}}
=\Gamma_{\bk\bkp}^{a,\mathrm{IV}}= U_{\bk-\bkp}^2\Pi^{A}_{\bk-\bkp}(\omega-\omega')=U_{\bk-\bkp}\bq^2 \left[\Pi^{R}_{\bk-\bkp}(\omega-\omega')\right]^* $
 in regions III and IV, where $\I(\omega-\omega')<0$.
Combining the contributions from regions II-IV, we obtain
\bwt
 \bea
 \Gamma^{\{a\}}_{\bk\bkp}
 =U_{\bk-\bkp}^2\left\{2\R \Pi^R_{\bk-\bk'}(\omega-\omega')\left[n_F(\omega')-n_F(\omega'+\Omega)\right]+2i\I \Pi^R_{\bk-\bk'}(\omega-\omega')\left[2n_B(\omega'-\omega)+n_F(\omega')+n_F(\omega'+\Omega)\right]\right\},\notag\\
 \label{vert3}
  \eea
\ewt
where $n_B(\epsilon)$ is a Bose function.

As before, we replace each of the two momentum integrals in Eq.~(\ref{vert2}) by integrals over the Fermi surface
and
over the dispersion,
and set $\bk=\bk_F$ and $\bkp=\bkp_F$ everywhere
except for the Green's functions
In the high-frequency regime, the Green's functions in Eq.~(\ref{vert2}) can be replaced by the
 bare ones [Eq.~(\ref{bare})]; then the product $\int d\ek G^R_{\bk}(\omega+\Omega)G^A_{\bk}(\omega)\int d\ekp G^R_{\bkp}(\omega'+\Omega)G^A_{\bk}(\omega')=-4\pi^2Z_{\bk_F}^2Z^2_{\bkp_F}/\Omega^2$  is real. Therefore, the imaginary part of the current-current correlator is given by $\I \Gamma^{\{a\}}_{\bk\bkp}$ from Eq.~(\ref{vert3}). Recalling that $\I\Pi_{\bq}(\Omega)=-{\cal D}_{\bq}\Omega$, where
\bea
D_{\bq}&=&\frac{1}{(2\pi)^{2}}\oint \frac{dA_{\bkp_F}}{v_{\bkp_F}} Z_{\bkp_F}Z_{\bkp_F+\bq}\delta(\epsilon_{\bkp_F+\bq})\left\vert_{\ekp=0}\right.\notag\\
\eea
(cf.  Eq.~(2.7) of I), and relabeling
 $\omega'\to \omega+\Omega'$, we obtain the contribution of diagram $a$ to the conductivity
 \bea
 \R\sigma_2^{\{a\}}
& -=&
 \frac{8\pi^2e^2
}{(2\pi)^D}\oint dA_{\bk_F}\oint dA_{\bkp_F}\int d\omega\int d\Omega' {\cal N}(\omega,\Omega',\Omega)
\notag\\
&&\times\frac{\bu_{\bk_F}\cdot\bu_{\bkp_F}}{v_{\bk_F}v_{\bkp_F}}Z_{\bk_F}^2Z^2_{\bkp_F}{\cal D}_{\bk_F-\bkp_F}U^2_{\bk_F-\bkp_F},
\label{sigma2a}
 \eea
where
\bea
{\cal N}(\omega,\Omega',\Omega)&=&\frac{\Omega'}{\Omega^3}\left[n_F(\omega)-n_F(\omega+\Omega)\right]\label{n}
\\&&\times \left[2n_B(\Omega')+n_F(\omega+\Omega')+n_F(\omega+\Omega'+\Omega)\right].\notag
\eea
The integral over $\Omega'$ in Eq.~(\ref{sigma2a}) is the same as in the
sum of the imaginary parts of the self-energies corresponding to diagram $a$ in Fig.~1 of I
[cf. Eq.~(2.5a) in I].
This integral gives $\frac{1}{2}\left(\omega^2+(\omega+\Omega)^2+2\pi^2T^2\right)$.
Averaging the last result with the difference of the Fermi functions, as in Eq.~(\ref{4pi2}),
we obtain
\be
\int d\omega \int d\Omega'{\cal N}(\omega,\Omega',\Omega)=\frac{1}{3}\frac{\Omega^2+4\pi^2 T^2}{\Omega^2}.\label{cfl_n}
\ee
Thus $\R\sigma_2^{\{a\}}$ differs from $\R\sigma_1$ in Eq.~(\ref{sigma1}) only by a prefactor, which can be read off from Eq.~(\ref{sigma2a}).
  \begin{figure}[t]
\includegraphics[width=0.3\textwidth]{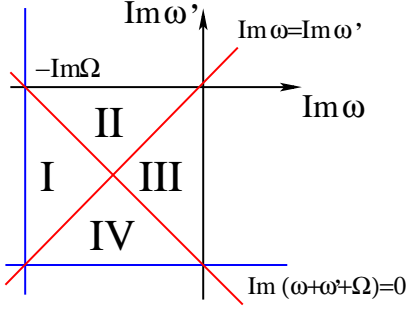}
\caption{(color on-line). Regions
of the $(\I\omega,\I\omega')$ plane.}
\label{fig:regions}
\end{figure}

Other diagrams for ${\cal K}^R_2(\Omega,T)$ can be analyzed in a similar fashion. For example, diagrams $b$ and $c$ in Fig.~\ref{fig:vertex} are similar to diagram $a$ with the only difference that $U_{\bk-\bk'}^2\I\Pi^R_{\bk-\bkp}(\omega-\omega')$ in Eq.~(\ref{vert3}) is replaced by $U_{\bk-\bkp}\I{\cal P}^R_{\bk,\bkp}(\omega'-\omega)$, where
\bea
&&\I{\cal P}^R_{\bk,\bkp}(\Omega)=\int d\epsilon\left[n_F(\epsilon)-n_F(\epsilon+\Omega)\right]\notag\\
&&\times\int_{\bp}\I G_{\bp}(\epsilon)\I G_{\bp+\bkp-\bk}(\epsilon+\Omega)U_{\bk-\bp}.
\eea
In I, we showed that the analytic properties of ${\cal P}^R$ in the frequency plane are the same as those of the polarization bubble. Therefore, the contributions of diagrams $b$ and $c$ to $\R\sigma$, which are equal to each other, also scale
 as $(\Omega^2+4\pi^2T^2)/\Omega^2$ with a prefactor different  from that of diagram $a$. [For $U_\bq=\mathrm{const}$,
the combined contribution of $b$ and $c$ cancels that of $a$.]

The Cooper-channel vertex--diagram $d$--appears to be somewhat different from particle-hole diagrams $a$-$c$ but, in fact, it gives the same result. To see this, we notice that the Matsubara vertex in diagram $d$ depends only on the combination $\omega_m+\omega_{m'}+\Omega_n$; hence, analytic properties of the retarded vertex depend on whether one is above or below the  $\I\omega+\I\omega'+\I\Omega=0$ line in Fig.~\ref{fig:regions}. Therefore, $\Gamma^{\mathrm{II}}_{\bk\bkp}=\Gamma^{\mathrm{III}}_{\bk\bkp}=\left(\Gamma^{\mathrm{IV}}_{\bk\bkp}\right)^*$, and the vertex reduces to
\bwt
 \bea
 \Gamma^{\{d\}}_{\bk\bkp}
 =2\R {\cal C}^R_{\bk\bkp}(\omega,\omega',\Omega)
 \left[n_F(\omega')-n_F(\omega'+\Omega)\right]+2i\I {\cal C}^R_{\bk\bk'}(\omega,\omega',\Omega)\left[2n_B(\omega'+\omega+\Omega)+n_F(\omega')+n_F(\omega'+\Omega)\right].\notag\\
 \label{vert4}
  \eea
\ewt
As before, we need only the imaginary part of of the vertex which contains
\bea
\I {\cal C}^R_{\bk\bk'}&=&\int_{\bp}\int\frac{d\epsilon}{\pi}\I G^R_{\bp}(\epsilon)\I G^R_{\bk+\bkp-\bp}(\omega+\omega'+\Omega-\epsilon)\notag\\
&&\times\tanh\frac{\epsilon}{2T}U_{\bk-\bp}U_{\bp-\bkp}.
\eea
[For $U_{\bk}=\mathrm{const}$, ${\cal C}$ reduces to a Cooper bubble.] Substituting explicit expressions for the spectral functions and integrating over $\epsilon$ and $\epsilon_{\bp}$, we obtain $\I {\cal C}^R_{\bk\bk'}={\cal C}_0(\omega+\omega'+\Omega)$ in the low-frequency limit,
where ${\cal C}_0$ contains a product of two interactions averaged over the FS. Substituting this result into the imaginary part
of Eq.~(\ref{vert4}), and relabeling $\omega\to -\omega$ and $\omega+\omega'+\Omega\to \Omega'$, we again arrive at the same integral as in Eq.~(\ref{cfl_n}).

The recipe for extracting the $\Omega^2+4\pi ^2 T^2$ scaling form of $\R\rho$ from a diagram of arbitrary order is now clear: one needs to extract a factor of $\Omega$  from either
a particle-hole or particle-particle convolutions of the Green's functions
 in the vertex, integrate it with the combination of the Fermi and Bose functions in Eq.~(\ref{n}), and then average the result with the difference of the Fermi function.
Up to a prefactor, all diagrams produce the same $\Omega/T$ scaling form of $\R\rho$ given by Eq.~(\ref{gurzhi}).
As it was the case with the self-energy considered in I, the overall prefactor cannot be expressed in a compact form.

\subsubsection{Non-canonical Fermi liquids}
\label{sec:ncfl}

The analysis of the preceding section was limited to the case of a conventional FL, when the infrared singularity arising from the $\Omega/q$ scaling
of the polarization bubble is suppressed by the phase space volume, and each of the diagram considered above is convergent on its own. In $D\leq 2$, the phase space is too small to suppress the singularities and the self-energy scales with $\omega$ in a non-canonical way: as $\omega^2\ln|\omega|$ in $D=2$ and as $|\omega|^D$ in $1<D<2$.
However, infrared singularities in different diagrams for the conductivity cancel each other.
This cancelation manifests the gauge-invariance of the conductivity.
In perhaps more familiar terms, this effect makes the conductivity to depend on the transport rather than single-particle relaxation time. It is more convenient to see this effect
in the Boltzmann equation, where the electron-electron contribution to the conductivity is expressed via a change in the electron
current carried by two electrons before and after a collision:\cite{rosch05,maslov1} $\left(\Delta\bv\right)^2\equiv \langle \left(\bv_{\bk}+\bv_{\bp}-\bv_{\bk-\bq}-\bv_{\bp+\bq}\right)^2\rangle$, where $\langle\dots\rangle$ stands for averaging over the FS.   For $q\to 0$, $\left(\Delta\bv\right)^2$ vanishes as $q^2$, which suppresses the infrared singularity for $D>1$.

To see how the same cancelation occurs in the Kubo formula, we consider two diagrams: diagram 1 in Fig.~\ref{fig:kubo} with both Green's functions dressed by a single-bubble self-energy correction (diagram $a$ in Fig.~ 1 of I) and diagram 2 in Fig.~\ref{fig:kubo} with vertex correction $a$ in Fig.~\ref{fig:vertex}. As we
are
 interested in the $q=0$ limit, it is convenient to decompose the momentum transfer $\bq$ into components along and perpendicular to the local Fermi velocity: $\bq= q_{||}{\hat\bv}_\bk+\bq_{\perp}$, where ${\hat\bv}_{\bk}=\bv_{\bk}/v_{\bk}$,  $q_{||}\ll q_{\perp}\ll {\bar k}_F$, and ${\bar k}_F$ is the characteristic \lq\lq radius\rq\rq\/ of the FS.
Accordingly, the (renormalized) dispersion is expanded as
\be
\varepsilon_{\bk+\bq}=\ek+v_{\bk}q_{||}+q_\perp^2/2m_{\bk},
\ee
where $m_{\bk}$ measures the local curvature of the FS.
The imaginary part of the self-energy insertions into  diagram 1of Fig.~\ref{fig:kubo}) is given by [cf. I, Eq.~(2.5a)]
\bea
\I\Sigma^{R,a}_\bk(\omega,T)&=&\int_\bq U^2_{\bq}\int\frac{d\Omega}{\pi}\left[n_B(\Omega)+n_F(\omega+\Omega)\right]\notag\\
&&\times\I G^R_{\bk+\bq}(\omega+\Omega)\I\Pi^R_\bq(\Omega),\label{imsigma}
\eea
where, as before, $\I\Pi^R_\bq(\Omega)=-{\cal D}_{\bq}\Omega$ and ${\cal D}_{\bq}\propto 1/q$ at $q\to 0$.
We neglect $q_{||}$ everywhere but in $\I G^R_{\bk+\bq}(\omega+\Omega)$, integrate over $q_{||}$, and substitute the result into Eq.~(\ref{k1}), which is then expanded in $\I\Sigma^R$. This yields
\bea
\I{\cal K}^R_1(\Omega,T)&=&\frac{\Omega}{\pi D}\int \frac{dA_{\bk_F}}{(2\pi)^D}\frac{u^2_{\bk_F}}{v^2_{\bk_F}}Z_{\bk_F}^3\int d\omega\int d\Omega'{\cal N}(\omega,\Omega',\Omega)\notag\\
&&
\times
\int\frac{d^{D-1}q_\perp}{(2\pi)^{D-1}}
{\cal D}_{\bq_\perp}U_{\bq_{\perp}}^2Z_{\bk_F+\bq_\perp},
\label{6_24a}
\eea
where ${\cal N}$ is given by Eq.~(\ref{n}). In $D\leq 2$, the $\bq_\perp$ integral is Eq.~(\ref{6_24a} infrared divergent.  However, this divergence is canceled by vertex part $a$ in Fig.~\ref{fig:vertex}. To see this cancelation, we need to assume that not only the charge velocity, defined by Eq.~(\ref{uf}), but also its derivative on the FS is known. Then, relabeling $\bk'=\bk+\bq$ in Eq.~(\ref{vert2}), we expand $\bu_{\bk+\bq}$ as
\be
\bu_{\bk+\bq}=\bu_{\bk}+\left(\left[{\hat\bv}_\bk q_{||}+\bq_\perp\right]\cdot\boldsymbol{\nabla}\right)\bu_\bk.
\ee
Substituting this expansion into Eq.~(\ref{vert2}) with $\Gamma=\Gamma^{\{a\}}$, we obtain
\bwt
\bea
\I{\cal K}^R_{2a}(\Omega,T)&=&\frac{\Omega}{2\pi^2D}\int \frac{dA_{\bk_F}}{(2\pi)^Dv_{\bk_F}}\int d\ek\int \frac{dq_{||}} {2\pi}\int
\frac{d^{D-1}q_{\perp}}{(2\pi)^{D-1}} \int d\omega \int d\Omega' \bu_{\bk_F}\cdot \left(\bu_{\bk_F}+\left(\left[{\hat\bv}_{\bk} q_{||}+\bq_\perp\right]\cdot\boldsymbol{\nabla}\right)\bu_\bk\left\vert_{\bk=\bk_F}\right.\right)\notag\\
&&\times G^R_{\bk}(\omega+\Omega)G^A_{\bk}(\omega)G^R_{\bk+\bq}(\omega+\Omega'+\Omega)G^A_{\bk+\bq}(\omega+\Omega')
U^2_{\bq}{\cal D}_{\bq}{\cal N}(\omega,\Omega',\Omega).
\label{6_24c}
\eea
\ewt
First, we integrate the product $G^R_{\bk+q}G^A_{\bk+\bq}\bu_{\bk+\bq}$ over $q_{||}$, setting $q_{||}=0$ everywhere else in the integrand. The $q_{||}$ independent and linear-in-$q_{||}$ terms in $\bu_{\bk+\bq}$ produce two integrals
\begin{eqnarray}
&&\int \frac{dq_{||}}{2\pi }\left(
\begin{tabular}{l}
\ensuremath{1} \\
\ensuremath{q_{||}}
\end{tabular}
\right)G_{\mathbf{k+q}}^{R}\left(\omega'
+\Omega \right) G_{\mathbf{k+q}}^{A}\left( \omega'
\right)
=
\left(
\begin{tabular}{l}
\ensuremath{\frac{iZ^2_{\bk_F+\bq_\perp}}{v_{\mathbf{k}}\Omega }}\\
\ensuremath{\frac{Z^2_{\bk_F+\bq_\perp}}{iv_{\mathbf{k}}^2\Omega }\frac{q^2_\perp}{2m_\bk}}
\end{tabular}
\right),
\notag\\
\end{eqnarray}
where terms of order $\Omega$, $T$ were neglected compared to $q_{\perp}^2/2m_{\bk}$ in the second line.
Next, we integrate
$G^R_{\bk}G^A_{\bk}$ over $\ek$, setting $\ek=0$ everywhere else in the integrand.
 The term proportional to $\bq_\perp$ vanishes by symmetry,
 and
 we
 obtain
\bea
\I\mathcal{K} ^{R}_{2a}\left( \Omega \right)  &=&-\frac{\Omega}{\pi D}\oint \frac{dA_{\bk_F}}{(2\pi)^D}\int d\omega\int d\Omega'{\cal N}(\omega,\Omega',\Omega)\notag\\
&&\times\frac{{\bf u}_{\bk_F}\cdot{\bf w}_{\bk_F}}{v_\bk^2}Z^2_{\bk_F},
\label{6_24b}
\eea
where
\bea
\mathbf{w}_{\mathbf{k}_F}&=&\int \frac{
d^{D-1}q_{\perp }}{\left(2\pi\right)^{D-1}}
\left[{\bf u}_{\bk_F}-\frac{q_\perp^2}{2m_{\bk_F} v_{\bk_F}}\left({\hat\bv}_\bk\cdot\boldsymbol{\nabla}_\bk\right){\bf u}_\bk\left\vert_{\bk=\bk_F}\right.\right]\notag\\
&&\times{\cal D}_{\bq_\perp}U^2_{\bq_\perp}Z^2_{\bk_F+\bq_\perp}.
\eea
The sum of two contributions to $\I{\cal K}$, i.e., Eqs.~(\ref{6_24a}) and (\ref{6_24b}),
contains a combination
\bea
&&\int \frac{dA_{\bk_F}}{v_{\bk_F}^2}Z_{\bk_F}^2Z_{\bk_F+\bq}
\Big{[}u_{\bk_F}^2(Z_{\bk_F}-Z_{\bk_F+\bq_\perp})\\&&+\left.\frac{q_\perp^2}{2m_{\bk_F} v_{\bk_F}}\bu_{\bk_F}\cdot\left({\hat\bv}_\bk\cdot\boldsymbol{\nabla}_\bk\right){\bf u}_\bk\left\vert_{\bk=\bk_F}\right.Z_{\bk_F+\bq_\perp}\right],\notag
\eea
which vanishes as $q_\perp^2$ for $q_\perp\to 0$
 and thus suppresses the $1/q_\perp$ divergence for all $D>0$.  Therefore, the optical resistivity  in the high-frequency regime has the same $\Omega^2+4\pi ^2 T^2$ form both in conventional and non-conventional FLs.
Notice that the other transport coefficients behave differently in these two cases;
for example, the {\em dc} thermal conductivity of a 2D FL behaves as
$1/T
\ln T$, as opposed to the $1/T$ behavior in 3D.~\cite{lyakhov:03}
 \subsection{
Bosonic first-Matsubara-frequency rule}
 Just as it was the case for a single-particle self-energy considered in I, the $\Omega^2+4\pi ^2 T^2$ scaling form of the optical resistivity can be related to the analytic properties of the current-current correlator along the Matsubara axis.

First, we consider the Matsubara version of diagram 1 in Fig.~\ref{fig:kubo}, given by Eq.~(\ref{k1}).
Recalling that $\mathrm{sgn}\Sigma_{\bk_F}(\omega_m)=\mathrm{sgn}\omega_m$, we integrate over $\ek$ to obtain (for $\Omega_n>0$)
\bea
&&{\cal K}_1(\Omega_n,T)=-\frac{2iT}{D(2\pi)^{D-1}}\oint dA_{\bk_F}\frac{u_{\bk_F}^2}{v_{\bk_F}}\\&&\times \sum^{\omega_m=-\pi T}_{\omega_m=-\Omega_n+\pi T}\frac{1}{\frac{i\Omega_n}{Z_{\bk_F}}+\Sigma_{\bk_F}(\omega_m+\Omega_n)+\Sigma_{\bk_F}(\omega_m)}.\notag
\eea
For $\Omega_n=2\pi T$, only one term with $\omega_m=-\pi T$ survives in the sum
\bea
&&{\cal K}_1(2\pi T,T)=
-\frac{2iT}{D(2\pi)^{D-1}}\oint dA_{\bk_F}\frac{u_{\bk_F}^2}{v_{\bk_F}}\notag\\
&&\times\frac{1}{\frac{2\pi i T}{Z_{\bk_F}}+\Sigma_{\bk_F}(\pi T)+\Sigma_{\bk_F}(-\pi T)}.
\eea
Since the single-particle self-energy satisfies the first (fermionic) Matsubara frequency rule,
i.e.,
$\Sigma_{\bk_F}(\pi T)=\Sigma_{\bk_F}(-\pi T)=0+\mathcal{O}(T^D
)$, the residual interaction drops out from
${\cal K}_1(2\pi T,T)$
(to order $T^3$).
Consequently, $\I{\cal K}_1^R(\Omega,T)$ vanishes (again, up to $\mathcal{O}(T^D)$ terms],
 when continued to the first non-zero
bosonic Matsubara-frequency $2\pi iT$.

Next, we integrate over $\ek$ and $\ekp$ as over independent variables in the Matsubara version of the vertex part, Eq.~(\ref{16}),  setting $\ek=\ekp=0$ in the rest of the integrand. At $\Omega=2\pi T$, only the $\omega_{m}=\omega_{m'}=-\pi T$ terms survive in the fermionic Matsubara sums and, as before, the fermionic self-energies, evaluated at $\pm\pi T$, drop out. Therefore,
\bea
&&{\cal K}_2(2\pi T,T)=-\frac{2}{D(2\pi)^{D}}\oint dA_{\bk_F}\oint dA_{\bkp_F}\frac{\bu_{\bk_F}\cdot\bu_{\bkp_F} }{v_{\bk_F}v_{\bkp_F}}\notag\\&&\times Z_{\bk_F}^2Z_{\bkp_F}^2\Gamma_{\bk_F,\bkp_F}(-\pi T,-\pi T, 2\pi T).
\label{k2m}
\eea
Particle-hole diagrams for $\Gamma_{\bk,\bk}(\omega_m,\omega_{m'},\Omega_n)$, e.g., diagrams $a$-$c$ in Fig.~\ref{fig:vertex},  depend on $\omega_m-\omega_{m'}$, while
particle-particle diagrams, e.g., diagram $d$, depend on $\omega_m+\omega_m'+\Omega_n$. Since both combinations of the frequencies vanish at $\omega_m=\omega_{m'}=-\pi T$ and $\Omega_n=2\pi T$, the vertex in Eq.~(\ref{k2m})
is {\em static}, i.e., it does not contribute to the real part of the conductivity. Therefore, the vertex-part contribution to
$\I{\cal K}^R(\Omega,T)$ vanishes as $\Omega=2\pi i T$ as well.

Strictly speaking, the proof presented above is valid only for conventional FLs, because
 the single-particle
self-energy
 obeys  the fermionic
first-Matsubara-frequency rule
 only in this case.
  However, as it was the case in Sec.~\ref{sec:ncfl}, deviations from the canonical behavior caused by infrared singularities must cancel between different diagrams. We did not attempt to repeat the proof for non-conventional FLs because the final result, applicable to both conventional and non-conventional FLs, clearly shows that $\R\sigma$ vanishes at $\Omega=2i\pi T$.

\section{Kubo formula:\\ zero-bubble approximation}
\label{sec:4}

\subsection{High-frequency regime}
\label{sec:drude}
Having shown that all diagrams for the conductivity produce the same scaling form in the high-frequency regime, we now
consider the case of lower frequencies, when $\Omega\lesssim Z_{\bk_F}\Sigma_{\bk_F}(\Omega,T)$.
The full analysis of the Kubo formula in this regime is rather involved.  We will simplify our task and focus on diagram 1 in Fig.~\ref{fig:kubo} which does not include vertex corrections.
Although such an approximation can be rigorously justified only in a few special cases, e.g.,
 in the $D=\infty$ limit of the Hubbard model,\cite{georges:96} or
 when electron-electron scattering connects points of FS with mutually perpendicular Fermi velocities,\cite{tsvelik}
  it provides a convenient
way to describe a crossover between high- and low-frequency regimes.
 We also
   adopt a slightly different version of the FL theory, compared to that considered in the preceding part of the paper. Namely, we assume that the self-energy
 is isotropic and local, i.e., that
it depends on $\omega$ much stronger than on $k-k_F$,
 and include the effects of interactions at all energy scales into the self-energy
 which, to order $\omega^2, T^2$ is
 now
 given by
         $\Sigma^R(\omega,T)=\omega(1+\lambda)+iC\left(\omega^2+\pi^2T^2\right)$.
         (Since we neglected the variation of the self-energy over the Fermi surface, the subscript $\bk_F$ will be suppressed from now on.)
        Since this model
        accounts for effects on interaction at all energy scales,
    the charge and Fermi velocities
      entering
      the conductivity diagram coincide with the bare Fermi velocity: $\bu_{\bk_F}=\bv_{\bk_F}=\bv^0_{\bk_F}$. Integrating over the bare dispersion, $\ek^0$, we obtain for the conductivity given by diagram 1
  \be
\sigma_1(\Omega,T)= \frac{i\omega_{p0}^2}{4\pi\Omega} \int_{-\infty}^\infty  d \omega \frac{n_F (\omega- \Omega) - n_F (\omega)}{\Omega + \Sigma^R(\omega,T) + \Sigma^R(\Omega - \omega, T)},
 \label{17}
 \ee
 where the bare plasma frequency is given by
 \be
 \frac{\omega_{p0}^2}{4\pi}=\frac{2e^2}{D(2\pi)^ D}\oint dA_{\bk_F}v^0_{\bk_F}.
 \ee
To cast the high-frequency limit of Eq.~(\ref{17}) into a form of the \lq\lq extended Drude formula\rq\rq\/,~\cite{basov:11}
we expand in $\I\Sigma^R$, evaluate the frequency integral, and bring the result
of integration back into the denominator,
which yields
  \be
 \sigma_1^{\mathrm{HF}}(\Omega,T) = i \frac{\omega^2_{p0}}{4\pi } \frac{1}{\Omega (1 + \lambda) + i \frac{2C}{3} \left(\Omega^2 + 4 \pi^2 T^2\right)}
 \label{19}
 \ee
 or
 \be
 \R\rho^{\mathrm{HF}}_1 (\Omega, T) =  \frac{4\pi}{\omega^2_{p0}}\frac{2C}{3}\left(\Omega^2 + 4 \pi^2 T^2\right),
 \label{20}
 \ee
where HF stands for \lq\lq high frequency\rq\rq\/.
 We remind that Eqs.~(\ref{19}) and (\ref{20}) are still valid only the high-frequency limit, defined by Eq.~(\ref{hf}).\\

\subsection{Low-frequency regime}
\label{sec:3}

 We now analyze $\sigma_1(\Omega,T)$
 at $\Omega \to 0$ and  finite $T$, when condition (\ref{hf})
is no longer valid.

In what follows, we will need the numerical values of integrals
\be
I_n=\int^{\infty}_0\frac{dx}{\cosh^2x}\frac{1}{\left(x^2+\pi^2/4\right)^{n+1}},
\ee
in particular, $I_0=0.333\dots$, $I_1=0.117\dots$, $I_2=0.043\dots$
 Substituting $\Omega=0$ into Eq.~(\ref{17}) and inegrating over $\omega$,
 we obtain the {\em dc} resistivity as
\be
\rho_1^{\mathrm{LF}} (0,T)=
\frac{4\pi}{\omega^2_{p0}} \left(a_0 C\right)  4\pi^2 T^2,
\label{dc1}
\ee
where LF stands for \lq\lq low frequency\rq\rq\/ and
\be
a_0=\frac{2}{I_0\pi^2} = 0.608\dots
\label{dc1_1}
\ee
On the other hand, extrapolation of the high-frequency conductivity in Eq.~(\ref{19}) to $\Omega=0$ gives
\be
\rho_1^{\mathrm{HF}} (\Omega\to 0, T)
  =\frac{4\pi}{\omega_{p0}^2} \frac{2C}{3} 4\pi^2 T^2
\label{dc_1}
\ee
We see that, in the zero-bubble approximation, the prefactors
in Eqs.~(\ref{20}) and (\ref{dc_1}) turn out to be very close to each other: $(2/3)/0.608= 1.097$.

We can also obtain the frequency dependence of $\sigma_1(\Omega,T)$ at $\Omega\to 0$
by expanding Eq.~(\ref{17}) further in $\Omega$
and casting the result into the form of Eq.~(\ref{19}). Expanding in $\Omega$ and evaluating the integrals over $\omega$, we obtain
\bwt
\be
\sigma_1^{\mathrm{LF}}(\Omega,T) =i \frac{\omega_{p0}^2}{4\pi} \frac{1}{a_1 \Omega (1 + \lambda) +
ia_0C \left(4\pi^2 T^2 + a_2\Omega^2 +
ia_3 \frac{(1+\lambda)^2\Omega^2}{C^2T^2}\right)},
\label{21_a}
\ee
or
\be
\rho_1^{\mathrm{LF}} (\Omega,T) = \frac{4\pi}{\omega^2_{p0}} a_0 C \left(4\pi^2 T^2 + a_2\Omega^2 + a_3 \frac{(1+\lambda)^2\Omega^2}{C^2T^2}\right).
\label{21_b}
\ee
In Eqs.~(\ref{21_a}) and (\ref{21_b}), $a_0$ is the same as in  Eq.~(\ref{dc1_1}) while
\bea
a_1&=&a_0^2I_1\pi^4/4=1.053\dots\notag\\
a_2&=&=\frac{a_0\pi^4}{4}\int^\infty_0\frac{dx}{\cosh^2x}\frac{1}{\left(x^2+\frac{\pi^2}{4}\right)^3}
\left[\left(x^2+\frac{\pi^2}{4}\right)
\left\{1-2x\tanh x+\frac{2}{3}\frac{1-2\sinh^2(x)}{\cosh^2x}\right\}\left(x^2+\frac{\pi^2}{4}\right)-2x^2\right]=1.030\dots\notag\\
a_3&=&\frac{\pi^4a_0}{32}\left(I_2-\frac{\pi^2 a_0}{2}I_1^2\right)=0.0036\dots
\eea
\ewt
In the FL regime, the imaginary part of the self-energy, $\sim CT^2$, must be much smaller than $T$. Therefore,
an expansion in $\Omega$ should be in powers of $\Omega/CT^2$. This is how the first and the last terms (with coefficients $a_1$ and $a_3$, correspondingly) in the denominator of Eq.~(\ref{21_a}) were obtained. However, because $a_3$ happens to be numerically very small, we also included the leading term from the expansion in $\Omega/T$ (with coefficient $a_2$).
In practice, the last term in Eq.~(\ref{21_a}) can be ignored so that
\be
\R\rho_1^{\mathrm{LF}} (\Omega,T) \approx \frac{4\pi}{\omega^2_{p0}}  a_0 C \left(4\pi^2 T^2 + a_2 \Omega^2\right),\ee
which is
 again
very close to the high-frequency form, Eq. (\ref{20}).

 Nevertheless, a change in the ratio of the  prefactors in the $\Omega^2$ and $T^2$ terms between the low- and high-frequency regimes
 indicates that the actual dependence of $\R\rho_1(\Omega,T)$ is actually more complex than just a sum of the $\Omega^2$ and $T^2$ terms.
We computed $\sigma_1(\Omega,T)$ numerically
and found that the $\Omega$ and $T$ dependences
 of $\R\rho_1 (\Omega,T)$ in the entire range $\Omega,T\ll E_F$ are
 well described by
 an approximate relation
 \be
 \R\rho_1(\Omega,T) = \frac{4\pi}{\omega^2_{p0}} \frac{2C}{3} \left(\Omega^2 + 3.65 \pi^2 T^2\right).
 \label{22}
 \ee
We see that
 the ratio of the $\pi^2 T^2$ and $\Omega^2$ terms
in $\R\rho_1(\Omega, T)$ is not equal
 to $4$,
but
numerically is quite close  to $4$.
 Notice, however, that a remarkable agreement between the low- and high frequency limits is valid only within
the zero-bubble approximation. We discuss effects not captured by
 this
 approximation in Sec.~\ref{sec:5}.

\subsection
{Incoherent regime}
\label{sec:inc}

Equation~(\ref{20}) is valid in the high-frequency regime, as specified by Eq.~(\ref{hf}).
Such a regime always exist in a coherent FL, where $\I\Sigma^R_{\bk_F}(\Omega,T)\ll\max\{\Omega,T\}$.
However, the optical conductivity of strongly correlated metals is often measured in the incoherent regime
 where
  all energy scales are comparable, i.e.,
$\Omega\sim T\sim \R \Sigma^R\sim \I\Sigma^R$.
 Having this in mind,
 it is instructive
 to study the behavior of $\R\rho_1(\Omega,T)$ in the incoherent regime.
 In general, calculations in this regime require a detailed knowledge of the electron-electron interaction at all energy scales.
 We use here a simple
 model
 in which $\mathrm{Im} \Sigma^R (\omega,T)$ is assumed to follow
 the FL form $
 C\left(\omega^2 + \pi^2 T^2 \right)$ all the way up to some cutoff
 frequency $\Lambda$ and to vanish
 at larger frequencies. The KK transformation then yields
\bea
 \mathrm{Re} \Sigma^R(\omega,T) &=&
\frac{2C \Lambda \omega}{\pi} -
\frac{C}{\pi} \left(\omega^2 + \pi^2 T^2 \right)
\ln{\frac{\Lambda + \omega}{|\Lambda - \omega|}} \notag\\
&&\approx
 \frac{\lambda\omega}{1 + \frac{\omega^2 + \pi^2 T^2}{\Lambda^2}}.
\label{23}
\eea
where $\lambda = 2C \Lambda/\pi$. At the last step, we replaced the actual  $\mathrm{Re} \Sigma^R(\omega,T) $ by an interpolation formula which describes the limits of both small and large (compared to $\Lambda$) frequencies but does not have a kink at $\omega=\Lambda$.
 These  forms of $\R\Sigma^R$ and $\I\Sigma^R$  are substituted into the Kubo formula for the conductivity, Eq. (\ref{17}), and the integral over $\omega$ is calculated numerically.

We found that in a wide range of $\Omega$ and $T$, including $\Omega\sim T \sim \Lambda$,
the optical resistivity can be well approximated by
\be
\R\rho_1(\Omega,T) \approx
\frac{4\pi}{\omega_{p0}^2}0.64B \left(\Omega^2 + 3.86 \pi^2 T^2\right).
\label{25}
\ee
We see the same trend as we found earlier
in the low-frequency regime:
the $\Omega$ and $T$ dependencies of the optical
resistivity are well
 approximated by the $\Omega^2$ and $T^2$
 forms, although the actual function is more complex
 than just the sum of these two terms, and the ratio of the $\pi^2T^2$ and $\Omega^2$ terms is smaller than $4$
 but
 not far from $4$.

\section{
Umklapp processes}
\label{sec:5}
In Sec.~\ref{sec:kubo}, we showed that any diagram for the conductivity produces the same $\Omega/T$ scaling form as indicated in Eq.~(\ref{gurzhi}), with a prefactor which depends on the electron spectrum.  Since no restrictions were imposed on the change in the electron quasimomentum due to the interaction,
both the normal and Umklapp processes were implicitly taken into account.
The interplay between these two types of processes is different, however, in different frequency regimes.

 In the high-frequency regime, as specified by Eq.~(\ref{hf}), the resistivity is finite already in the presence of only  normal processes, provided that Galilean invariance is broken by
a
 lattice. Even on a lattice, however,
 the leading, $\Omega^2+4\pi^2 T^2$ term  vanishes in several special cases, e.g., for a
 quadratic {\em or} isotropic FS in any D, and for a convex {\em and} simply-connected in 2D.~\cite{gurzhi_ee,rosch05,maslov1}
 In these cases, the optical resistivity scales as $\max\{\Omega^4,T^4\}$.
  [The case of an isotropic {\em and} quadratic spectrum corresponds to a Galilean-invariant FL,
 the conductivity of which retains a free-electron Drude form
 regardless of
  the electron-electron interaction.]
 In what follows, we assume that the FS does {\em not} belong to any of the types specified above,
so that normal processes do contribute to the
 leading term in the
 resistivity.  If Umklapp processes are also allowed, they affect the resistivity as well.
  The prefactor $A'$ in Eq.~(\ref{gurzhi}) is proportional to the interaction vertex, $\Gamma$.
 In the high-frequency regime, $\Gamma$ is just a sum of the vertices for normal and Umklapp processes ($\Gamma_{\mathrm{N}}$ and $\Gamma_{\mathrm{U}}$, correspondingly), i.e.,
\be A'\propto  \Gamma'=\Gamma_{\mathrm{N}}+\Gamma_{\mathrm{U}}.\label{gamma_inf}\ee

In the opposite limit of $\Omega=0$, the resistivity of an impurity-free system is non-zero only in the presence of Umklapp scattering. However, once Umklapp processes are allowed, normal processes contribute as well,~\cite{maebashi97_98} at least as a correction to the Umklapp contribution (again, if the FS is not of one of the types specified in the preceding paragraph). The effective vertex $\Gamma$, entering Eq.~(\ref{dca}), is now a non-trivial function of $\Gamma_{\mathrm{N}}$ and $\Gamma_{\mathrm{U}}$ which can be represented in the following scaling form
\be
A\propto\Gamma=\Gamma_{\mathrm{U}}\Phi(\Gamma_{\mathrm{N}}/\Gamma_{\mathrm{U}}).
\label{gamma0}\ee
On general grounds, one can infer that $\Phi(x\to 0)=C_1+\mathcal{O}(x)$ and $\Phi(x\to \infty)=C_2$,
where $C_{1,2}$ are constants. The ratio $\Gamma_N/\Gamma_U$ and the function $\Phi(x)$ itself depend on
the details of
both the
 bandstructure
  and
  the
  interaction
and are by no means universal.
Therefore, prefactors $A'$ and $A$ differ by a non-universal factor, which is expected to be of order
  one but not specifically close to $1$.

Even if, for some reason, normal processes are absent, $A$ and $A'$ still differ because, when calculating the optical resistivity in the high-frequency regime, one expands the Green's functions in the self-energy and averages the result with the difference of the Fermi functions, while in the low-frequency regime the self-energy must be kept in the denominators of the Green's functions. Although it turns out that $A$ and $A'$  almost coincide in the zero-bubble approximation (cf. Sec.~\ref{sec:drude}), there is no guarantee that this remains true if vertex corrections are taken into account.

We conclude this section with a remark in regard to a statement by Rosch and Howell,~\cite{rosch05} who argued that the coefficients $\alpha_0$ and $\beta_0$ in $\R\rho(\Omega,T)=\alpha_0\Omega^2+\beta_0 T^2$ are not, in general, related. For reasons explained above, this statement is correct
   if  $\R\rho(\Omega,T)$ is supposed to describe the whole range of frequencies: from low to high.
 However, as we have already emphasized, the formula $\alpha_0\Omega^2+\beta_0 T^2$ with constant $\alpha_0$ and $\beta_0$ does not describe a crossover between the high- and low-frequency regimes, Nevertheless,
 $\alpha$ and $\beta$ are universally related
in the {\em high-frequency regime}, where
$\beta_0/\alpha_0=4\pi ^2$.

This section concludes our analysis of the conductivity of a FL.
 To summarize, we have shown that the scaling form of the optical resistivity in Eq.~(\ref{gurzhi}) is quite robust.
 In the high-frequency regime, this form is produced by all diagrams for the conductivity. If vertex corrections are neglected,
 then one can go beyond the high-frequency regime. It turns out that,
 with only small changes in the numerical coefficients, Eq.~ (\ref{gurzhi}) form works well beyond its nominal region of validity, i.e., both near the {\em dc} limit and at such high $\Omega$ and $T$ that the FL picture itself is not applicable.
 In other words, if the $\Omega$ and $T$ dependencies of the resistivity are determined by the electron-electron interaction,
 it is impossible to avoid the FL scaling form with a coefficient of the $T^2$ term either equal or very close to $4\pi^2$. As discussed in
 the
 next Section,
  this is {\em not} what the experiment shows.

\section{Comparison to experiment}
\label{sec:6}
\subsection{Summary of experimental observations:
Disagreement with the Fermi-liquid theory}
\label{sec:status}

Now we turn to the discussion of the existing experimental data on the $\Omega/T$ scaling of the optical resistivity.
Although the $\Omega^2$ dependence of $\R\rho(\Omega,T)$ was convincingly demonstrated in \lq\lq weakly correlated metals\rq\rq\/ (Au,Ag, and Cu),~\cite{christy}
the $T$ dependence, if measured, was found to result from the electron-phonon rather than the electron-electron interaction.
This is not surprising since the electron-electron interaction in these metals is relatively weak and one needs to go to very low temperatures to observe the $T^2$ dependence. To the best of our knowledge, the $\Omega^2+4\pi^2T^2$ scaling still has not been verified in weakly correlated metals.

On the other hand, the $\Omega/T$ scaling of the optical conductivity in strongly correlated metals has been studied quite extensively; a detailed summary of experimental observations can be found in Ref.~\onlinecite{timusk11}. The conclusion of these studies is quite surprising: fitting the measured
\lq\lq optical
scattering rate\rq\rq\/
 into a phenomenological form
\be
\frac{1}{\tau(\Omega,T)} 
\equiv
\frac{4\pi}{\omega_p^2}\R\rho(\Omega,T)
=\mathrm{const}\times\left(\Omega^2+b\pi^2 T^2\right)
\label{tau_exp}
\ee
 has {\em not} produced $b$ close to $4$ in {\em
 any}
 of the
 cases studied so far.
 In some cases, e.g., in the heavy-fermion compound URu$_2$Si$_2$~\cite{timusk11}
above the $17.5$ K  transition
into the \lq\lq hidden-order\rq\rq\/ (HO)
state
 and in
 rare-earth based
 doped Mott insulators (Ce$_{0.095}$Ca$_{0.05}$TiO$_{3.04}$\cite{tio3} and Nd$_{0.905}$TiO$_3$\cite{ndtio3}),
 $b$ has been found to be close to $1$ rather than to $4$.
 On the other hand, a recent study\cite{mirzaei}
 of the underdoped cuprate HgBa$_2$CuO$_{4+\delta}$ reports $b\approx 2.3$,
 while another study of an organic material from the BEDT-TTF  family
 reports $b\approx 5.6$.\cite{dressel}

In the preceding sections, we showed that $b\approx 4$ is a robust property of FLs with electron-electron interaction.
We must then conclude that even
though $1/\tau(\Omega,T)$ in
 the
 compounds mentioned above exhibits FL-like dependences on $\Omega$ and $T$, the lack of a
FL-like  $\Omega/T$ scaling indicates that
 these dependencies
do not come only
 from
the electron-electron interaction.
 In the remainder of this Section, we attempt to
 explain
 the discrepancy between the FL theory and
 the experiment.

\subsection{Elastic vs inelastic contributions to the single-particle self-energy}
\label{sec:elastic}

In this Section,
 we try to identify a mechanism responsible for
deviation of
 the observed coefficient $b$
from
 the FL value of $4$.
 In the preceding Sections, we analyzed the conductivity of a FL under an implicit {\em assumption} that
 the only scattering mechanism is the electron-electron interaction among
itinerant electrons.
  However, the FL of itinerant electrons
is not the only example of a FL. Another example is
 a FL state formed around magnetic impurities at energies below the Kondo temperature.
In the Kondo case, there are two channels of interaction:
an
elastic one, which contributes an $\omega^2$ term to the imaginary part of the self-energy, and
 an
 inelastic
 or electron-electron
  one,
which contributes an $\omega^2+\pi^2T^2$ term. The relative weight of these two contributions depends on the strength of the on-site electron-electron interaction, which can be conveniently parameterized by the Wilson ratio, $R$.~\cite{kondo}  In the unitary limit, when $R=2$, the elastic channel is twice more efficient than the inelastic one, i.e.,
\bea
\I\Sigma^R(\omega,T)&=&B-\frac{2}{3}C'
\left(\omega^2+\frac{1}{2}\left[\omega^2+\pi^2T^2\right]\right)\notag\\&&=B-C'\left(\omega^2+\frac{1}{3}\pi^2T^2\right),\label{sek}
\eea
where $B$ is the $\omega$-independent part of the elastic contribution and $C'>0$.  The reduction of the $T^2$ contribution
to $\I\Sigma_R$
is reflected in the 
optical scattering rate,
which is obtained, as before, by substituting  Eq.~(\ref{sek}) into the Kubo formula (\ref{17}) (in the zero-bubble approximation) and integrating over $\omega$:
\be
\frac{1}{\tau(\Omega,T)}
=B-\frac{2C'}{3}\left(\Omega^2+2\pi ^2T^2\right)
\label{tauk}
\ee
 Thus the Kondo FL belongs to a different universality class with $b=2$. This does not explain the experiment yet
because of the {\em
 non-metallic} signs of the $\Omega$ and $T$ dependences of $1/\tau(\Omega,T)$ in Eq.~(\ref{tauk}), as opposed to the {\em metallic}
 signs
 observed in the experiment at least at the lowest frequencies.
However, this gives us an idea to ask:
 how does a reduction of the inelastic contribution to the self-energy affect the relative weight of the $\Omega^2$ and $T^2$
terms in the optical conductivity?

To answer this question, we introduce a phenomenological form of the self-energy
\be
 \I\Sigma^R(\omega,T)=\Sigma_{\mathrm{el}}(\omega)+C\left(\omega^2+\pi^2T^2\right).
 \label{pheno_se_2}
 \ee
 The first term describes a contribution of the elastic channel which arises from the energy dependence of the effective scattering cross-section. However, since scattering is elastic, its cross-section does not depend on the temperature (provided that the number and other properties of the scattering centers do not vary with $T$) and
$\Sigma_{\mathrm{el}}(\omega)$  is $T$-independent.
The second term describes the contribution of inelastic electron-electron interaction,
 which is the same
 as in a conventional FL.
A particular form of $\Sigma_{\mathrm{el}}(\omega)$ is important for determining the actual behavior of the optical conductivity, especially if $\Sigma_{\mathrm{el}}(\omega)$ is a sharp function of $\omega$, as it is the case for resonant scattering, considered in the next Section. For the time being, however, we assume only
that $\Sigma_{{\mathrm el}}(\omega)$ is an analytic function of $\omega$ and expand it to second order in $\omega$ as
\be
\Sigma_{\mathrm {el}}(\omega)=\Sigma_{{\mathrm el}}(0)+\Sigma'_{{\mathrm el}}(0)\omega+a C\omega^2,
\label{taylor}
\ee
where the constant $C$ [the same as in Eq.~(\ref{pheno_se_2}] was factored out for convenience, and $a$ is another constant
which can be of either sign. We call the elastic contribution \lq\lq metallic\rq\rq\/ if $a>0$ and \lq\lq non-metallic\rq\rq\/ if $a<0$. On the other hand,
the inelastic contribution is always metallic because $C>0$ (which is not the case for the Kondo model).
Combining Eqs.~(\ref{pheno_se_2}) and (\ref{taylor}), we obtain
\be
 \I\Sigma^R(\omega,T)=\Sigma_{{\mathrm el}}(0)+\Sigma'_{{\mathrm el}}(0)\omega+C\left[a\omega^2+\left(\omega^2+\pi^2T^2\right)\right].
 \label{pheno_se_3}
 \ee
 The $a\omega^2$ term mimics the
   $\omega^2$ dependence of the inelastic contribution but does not have its $T^2$ counterpart.
 We emphasize that the  $\omega$ and $T$ dependencies of the inelastic contribution should be consistent
  with the fermionic first-Matsubara-frequency rule, which stipulates that the inelastic term in Eq.~(\ref{pheno_se_2}) must vanish
  upon replacing $\omega$ by $\pm i\pi T$. This rule, which is obviously satisfied with our choice for the inelastic part,
  does not allow for changes in the relative weight of the $\omega^2$ and $T^2$ terms in this part.
   Next, we substitute Eq.~(\ref{pheno_se_2}) into Eq.~(\ref{17}), integrate over $\omega$, upon which the linear-in-$\omega$ term in $\I\Sigma^R(\omega,T)$
   vanishes,
 and obtain the optical scattering rate as
\be
\frac{1}{\tau(\Omega,T)}=\frac{1}{\tau_0}+\frac{2}{3}(a+1)C\left[\Omega^2+b\pi^2T^2\right]
\label{tau}
\ee
with
   \be
   b=\frac{a+4}{a+1}.
   \label{b_def}
   \ee
   The residual term, $1/\tau_{0}$, contains a contribution from static disorder (not considered explicitly here), and we absorbed $\Sigma_{\mathrm{el}}$ into this term as well.

  Now we discuss constraints imposed on the parameter $a$, and thus on $b$.
  For $a<-1$, the prefactor of the second term in Eq.~(\ref{tau}) is negative, i.e., the $\Omega$ and $T$ dependencies of $1/\tau(\Omega,T)$ are non-metallic. Since this does not correspond to any of the experiments, we discard this possibility.
  The special case of $a=-1$ corresponds to $1/\tau(\Omega,T)$ which depends only on $T$ but not on $\Omega$. Discarding this possibility as well,  we focus on the range
 $-1<a<\infty$, which corresponds to $1\leq b <\infty$.  The  FL value of $b=4$ is reproduced for $a=0$.  The opposite limit of $a=\infty$ (and thus $b=1$) corresponds to a purely elastic scattering mechanism.
   The range $1<b<4$ corresponds to a mixture of elastic and inelastic mechanisms with $a>0$, i.e., with  a metallic sign of the elastic
   contribution,
  whereas $b>4$ corresponds to a non-metallic elastic contribution with $-1<a<0$, although the $\Omega$ and $T$ dependences of $1/\tau(\Omega,T)$ in this case are still metallic.

  According to
  this classification scheme,
    the value of $b\approx 1$, reported in Refs.~\onlinecite{timusk11,tio3,ndtio3} for the U, Ce, and Nd-based compounds, indicates a purely elastic scattering mechanism ($a=\infty$).
    The value of $b\approx 2.3$ (and thus $a\approx 1.3$), reported in Ref.~\onlinecite{mirzaei} for the Hg-based underdoped cuprate,  points at a mixture of elastic and inelastic mechanisms with comparable weights, and with a metallic sign of the elastic contribution. Finally,
    $b\approx 5.6$ (and thus $a\approx -0.35$), reported in Ref.~\onlinecite{dressel} for the organic material, also corresponds to a mixture of the two mechanisms
    but
    with a non-metallic sign of the elastic contribution.

The deviation from the FL behavior is the most dramatic for the $b=1$ case, where it appears that the electron-electron interaction does not play any role. However, this conclusion would be incorrect.
 In the next Section, we discuss one example of a purely elastic scattering mechanism, i.e., scattering from resonant levels,
and apply this model to the \ho~data.
 We will see that, while the optical conductivity
can be explained by resonant-level scattering alone, an explanation of the $T$ dependence of the  {\em dc} resistivity requires invoking
 a sufficiently strong electron-electron
interaction.

\subsection{Scattering from resonant levels:
\\
the case of \ho
}
\label{sec:resimp}
\begin{figure}[t]
\includegraphics[width=0.5\textwidth]{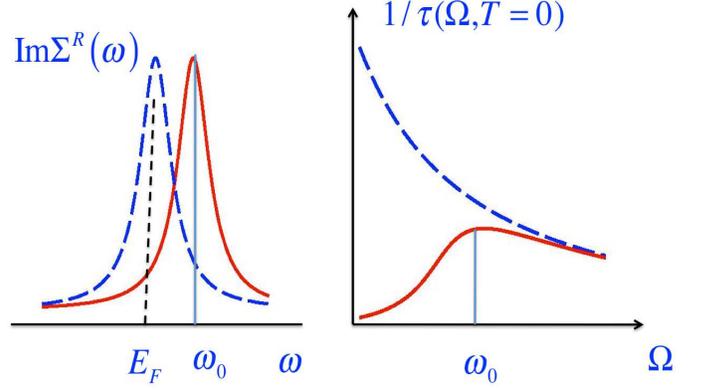}
\caption{(color on-line). Imaginary part of the
 fermionic
 self-energy, Eq.~(\ref{ser1}), (left) and optical
 scattering rate
 at $T=0$, Eq.~(\ref{ose1}) (right) for scattering at resonant impurities.}
\label{fig:resT=0}
\end{figure}

\begin{figure}[t]
\includegraphics[width=0.5\textwidth]{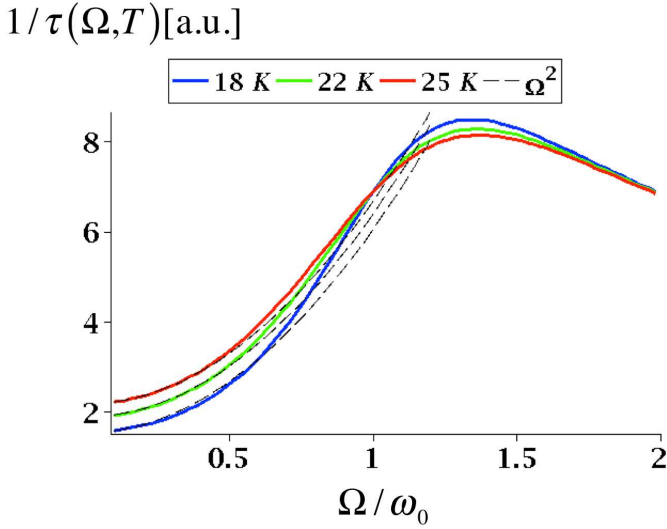}
\caption{(color on-line). Optical self-energy in the resonant-impurity model as a function of frequency at several temperatures. Absolute values of temperatures are fixed by choosing $\omega_0=12.5$ meV and $\gamma=0.2\omega_0$. Dashed lines show $\Omega^2$ fits of the actual dependencies.}
\label{fig:res_omega}
\end{figure}

\begin{figure}[t]
\includegraphics[width=0.3
\textwidth]{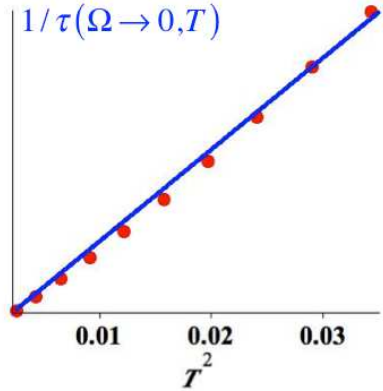}
\caption{(color on-line). The intercept, $1/\tau(\Omega\to 0,T)$, in the resonant impurity model as a function of $(T/\omega_0)^2$.}
\label{fig:intercept}
\end{figure}

\begin{figure}[t]
\includegraphics[width=0.5
\textwidth]{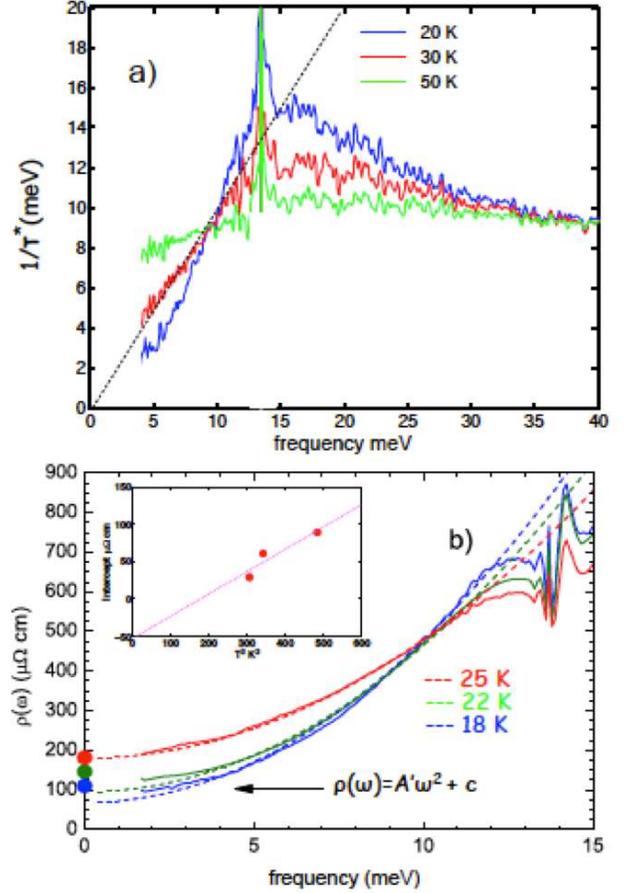}
\caption{(color on-line). Experimental results for in URu$_2$Si$_2$ from Ref.~\onlinecite{timusk11}.
a) Optical scattering rate $1/\tau(\Omega,T)$.
b) The optical resistivity at lower frequencies from the refined reflectivity.}
\label{fig:tom_data}
\end{figure}

In this section, we discuss
 the model of purely elastic scattering from resonant levels, located  at energy $\omega_0$ away from the Fermi energy and of width $\gamma$. The self-energy in this case is given by
\be
\I\Sigma^R(\omega,T)
=\Sigma_{\mathrm{el}}(\omega)
=\frac{C_0\gamma}{(\omega-\omega_0)^2+\gamma^2}.
\label{ser1}
\ee
At $T=0$, the corresponding optical scattering rate is given by
\be
\frac{1}{\tau(\Omega,0)}=\frac{C_0}{\Omega}\left[\arctan\frac{\Omega-\omega_0}{\gamma}+\arctan\frac{\Omega+\omega_0}{\gamma}\right].
\ee
If the resonant level coincides with the Fermi energy, $1/\tau(\Omega,0)$ is purely
 non-metallic, i.e., it decreases as $\Omega$ increases. If the resonant level is away from the Fermi energy, $1/\tau(\Omega,0)$ is a non-monotonic function of $\Omega$ with a maximum at $\Omega\sim\omega_0$ (see Fig.~\ref{fig:resT=0}). The origin of the maximum is clear: as $\Omega$ increases from zero to $\omega_0$, the rate of transitions from the Fermi energy to the resonant levels increases. When $\Omega$ becomes
 larger than $\omega_0$, the rate decreases because now the energy interval from the Fermi energy to the resonant level constitutes only a
  fraction of the photon energy.
Expanding Eq.~(\ref{ser1}) near $\omega=0$ as
 \be
\I\Sigma^R(\omega)=C_0\gamma\left[\frac{1}{\omega_0^2+\gamma^2}+\frac{2\omega\omega_0}{\left(\omega_0^2+\gamma^2\right)^2}
+\frac{3\omega_0^2-\gamma^2}{\left(\omega_0^2+\gamma^2\right)^3}\omega^2\right]
\label{ser}\ee
and substituting (\ref{ser}) into (\ref{17}), we obtain
\be
\frac{1}{\tau(\Omega,T)}=\mathrm{const}+\frac{C_0\gamma\left(3\omega_0^2-\gamma^2\right)}{\left(\omega_0^2+\gamma^2\right)^3}\left(\Omega^2+\pi^2T^2\right).
\label{ose1}
\ee
(The
linear in $\omega$  term in Eq.~(\ref{ser}) vanishes by parity.)
Already for a moderately narrow level, i.e., for $\gamma<\omega_0\sqrt{3}$, the signs of both the $\Omega$ and $T$ dependences of $1/\tau(\Omega,T)$ are metallic, and $b=1$.
The behavior of $1/\tau(\Omega,T)$ over a larger range of $\Omega$ and $T$ is obtained by substituting Eq.~(\ref{ser1}) into the Kubo formula (\ref{17}) and computing the integral over $\omega$ numerically. The results are shown in Fig.~\ref{fig:res_omega}. To compare to the experimental data on  URu$_2$Si$_2$ from Ref.~\onlinecite{timusk11}, shown in Fig.~\ref{fig:tom_data}, we choose $\omega_0=12.5$ meV to
 match the
 position of the peak in the data.
 All other energies are measured
 relative
 to $\omega_0$.
  In Fig.~\ref{fig:res_omega}, $\gamma=0.2\omega_0$, and the temperatures are chosen to coincide with the absolute temperatures used in the experiment ($18$, $22$, and $25$ K).
Comparing Figs.~\ref{fig:res_omega} and \ref{fig:tom_data}, we see that the model reproduces
 the characteristic features of the data, i.e.,
a non-monotonic dependence of $1/\tau(\Omega,T)$ on $\Omega$, as well as an
(approximate)
isosbestic
 point at $\Omega\approx \omega_0$, where $1/\tau(\Omega,T)$ apparently does not depend on $T$.
The very existence of the isosbestic point implies that the data cannot be described by Eq.~(\ref{gurzhi}) with a $T$-independent prefactor $A'$. Nevertheless, we follow the same protocol as used in Ref.~\onlinecite{timusk11}, i.e.,
 we
 fit
 the $\Omega$ dependence of
 the
 computed $1/\tau(\Omega,T)$ into
 an $\Omega^2$ function
 ( shown by dashed lines in Fig.~\ref{fig:res_omega}),
 then fit the intercept $1/\tau(\Omega\to0,T)$
 into
 a $T^2$ function (shown in Fig.~\ref{fig:intercept}), and take the ratio of the slopes of the $T^2$ and $\Omega^2$ fits. This procedure gives $b\approx 0.9$, which is within the margin of error of the experimental value $b=1\pm 0.1$.~\cite{timusk11}

 The behavior of $1/\tau(\Omega,T)$ in  the Ce and Nd compounds (Refs.~\onlinecite{tio3} and \onlinecite{ndtio3}, correspondingly) is qualitatively similar to that in \ho, although the ranges of $\Omega$ and $T$ are drastically different.
 In \nd, $1/\tau(\Omega,T)$ scales as $\Omega^2$ up to about $0.1$ eV, followed by a maximum at $\approx 0.27$\;eV. The $\Omega=0$  intercept of $1/\tau$ scales as $T^2$ over a wide temperature range: from $29$ to $295$ K.~\cite{slopes}
In \ce, $1/\tau(\Omega,T)$ scales as $\Omega^2$ also up to about $0.1$ eV, followed by a tendency to saturation; but the maximum is not yet revealed at the highest frequency measured ($\approx 0.14$ eV). The $\Omega=0$ intercept also scales as $T^2$ over a wide range of temperatures.
 These similarities suggest that, despite obvious differences in composition
and energy scales in U, Ce, and Nd compounds, the optical response in all three cases is governed by the same mechanism.

Elucidation of the microscopic mechanism of resonant levels is beyond the scope of this work,
and we make just a brief comment in this regard. It is very unlikely that clean samples studied in Refs.~\onlinecite{tio3,ndtio3,timusk11} contained considerable amounts of {\em extrinsic} resonant impurities. Therefore, resonant states must be {\em intrinsic} to these compounds. We surmise that $f$-electrons of U, Ce, and Nd atoms, although arranged into a sublattice, play the role of {\em incoherent} resonant levels at sufficiently high energy scales  probed in optical measurements.

 \begin{figure}[t]
\includegraphics[width=0.4
\textwidth]{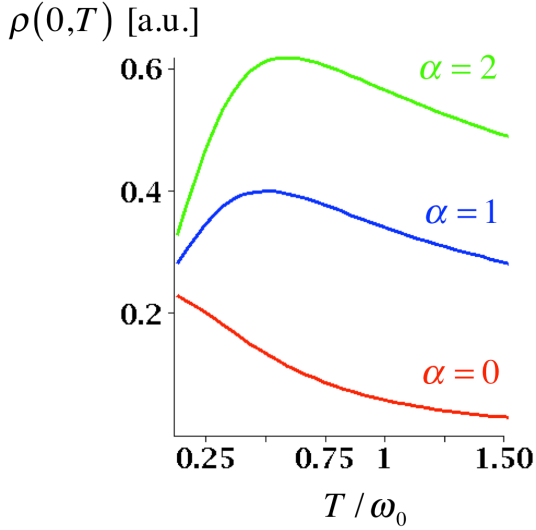}
\caption{(color on-line). {\em dc} resistivity (arbitrary units) for a model form of the self-energy which combines resonant-level and electron-electron contributions, Eq.~(\ref{comb}). Temperature is measured in units of the resonant-level energy, $\omega_0$, which
is also chosen to coincide with the cutoff energy $\Lambda$. The resonant-level width $\gamma=0.2\omega_0$. Parameter $\alpha$, defined by Eq.~(\ref{alpha}), measures the relative strength of the two contributions.}
\label{fig:dcrho}
\end{figure}

 \begin{figure}[t]
\includegraphics[width=0.4
\textwidth]{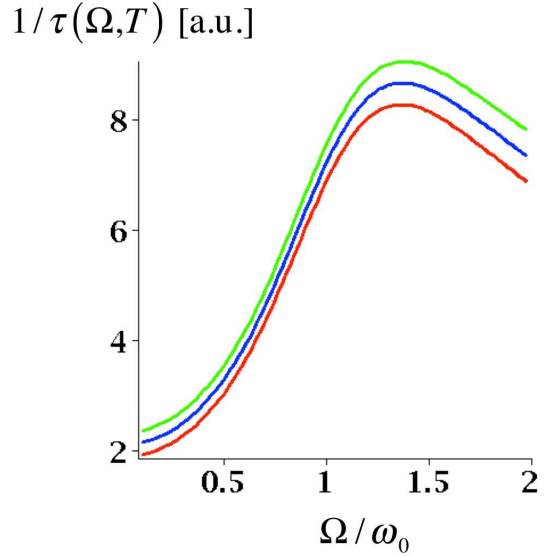}
\caption{(color on-line). Optical scattering rate (arbitrary units) for a model form of the self-energy which combines resonant-level and electron-electron contributions, Eq.~(\ref{comb}). Frequency is measured in units of $\omega_0$. $T=22$ K, $\gamma=0.2\omega_0$. From top to bottom: $\alpha=0$ (red), $\alpha=1$ (blue), $\alpha=2$ (green).}
\label{fig:opt_ri_ee}
\end{figure}

\subsection{Combined effect of the electron-electron and resonant-level scattering mechanisms}
\label{sec:combined}
Although the resonant-scattering model explains the results of optical measurements, this model alone cannot explain
the temperature dependence of the {\em} dc resistivity. Above the superconducting transition temperature ($\approx 1$ K) in \ho, both the $a$- and $c$-axis resistivities increase with $T$ in a quadratic manner within the HO phase, exhibit a kink at HO $T^{\mathrm{HO}}_c$, and continue to increase up to about $75$ K, where $\rho_{a}$ goes through a broad maximum
whereas
 $\rho_c$ starts to saturate.~\cite{palstra:86,mydosh:90,behnia:09}  The slopes of the increasing parts in $\rho_{a,c}$, both below and above $T^{\mathrm{HO}}_c$,
are largely independent of the
 residual resistivity,
\cite{timusk_private} which indicates that the $T$-dependence comes from
an intrinsic mechanism. On the contrary, the $T$ dependence of the {\em dc} resistivity in the resonant-scattering model is purely {\em
 non-metallic}.
Indeed, it is easy to see that the {\em dc} conductivity,
\be
\sigma(0,T)=\frac{\omega_{p0}^2}{8\pi}\int d\omega\left(-\frac{\partial n_F}{\partial\omega}\right)\frac{1}{\I\Sigma^R(\omega)}
\label{dc}\ee
with $\I\Sigma^R(\omega)$ from Eq.~(\ref{ser1}), increases with $T$ as $T^2$; therefore, $\rho(0,T)=1/\sigma(0,T)$ decreases with $T$. In order to reproduce the metallic sign of $\rho(0,T)$, at least for $T$ below $75$ K, one needs to bring in the inelastic electron-electron interaction with $\I\Sigma^R(\omega,T)$ given by Eq.~(\ref{cfl}). This seems to defy the purpose of the preceding analysis, as we have argued that the optical data cannot be explained by an inelastic mechanism. It turns out however, that a combination of elastic and inelastic mechanisms explains both the {\em dc} and optical data. In the \lq\lq combined\rq\rq\/ model, the total self-energy is a sum of two contributions
\bea
\I\Sigma^R(\omega,T)&=&\frac{C_0\gamma}{(\omega-\omega_0)^2+\gamma^2}\label{comb}\\
&&+C\left(\omega^2+\pi^2 T^2\right)F\left(\frac{\sqrt{\omega^2+\pi^2 T^2}}{\Lambda}\right).\notag
\eea
This equation is the same as we introduced in Eq.~(\ref{pheno_se_2}), except for
 now the electron-electron
contribution
contains a smooth cutoff function $F(x)$,
 defined in such a way that
$F(0)=1$ and
$F(x)$ falls off faster than $1/x^2$ for $x\gg 1$.
 The function $F(x)$ is chosen to reproduce a slow decrease of the measured $a$-axis resistivity at higher temperatures. (Since optical experiments probe
the basal-plane conductivity, we focus on this case.) To minimize the number of free parameters, we set $\Lambda=\omega_0$. The relative strength of two contributions to $\I\Sigma^R(\omega,T)$ in Eq.~(\ref{comb}) is controlled by a dimensionless parameter
\be
\alpha\equiv \frac{C\omega_0^4}{C_0\gamma}.
\label{alpha}\ee
Larger values of $\alpha$ correspond to a larger electron-electron and smaller resonant-level contribution, and vice versa.
Using the small-$\omega$ expansion in Eq.~(\ref{ser}), it is easy to show that $d\rho(0,T)/dT|_{T\to 0}$ is positive, i.e., \lq\lq metallic\rq\rq\/, already for $\alpha>1/4$. The dependence of $\rho(0,T)$ over the entire temperature range is obtained by numerical integration of Eq.~(\ref{dc}) with $\I\Sigma^R$ from Eq.~(\ref{comb}). The resulting profiles of $\rho(0,T)$ are shown in Fig.~\ref{fig:dcrho} for $\alpha=0,1,2$. As we see, the electron-electron contribution leads to a qualitative change in $\rho(0,T)$:
a purely
 non-metallic
 $T$ dependence with resonant levels alone ($\alpha=0$) is transformed into a curve with a maximum ($\alpha=1,2$). The $\alpha=2$ curve is already similar to the measured profile of $\rho(0,T$), which increases almost three-fold
when $T$ is varied in between  $T_c^{\mathrm{HO}}$ (chosen as the lowest temperature in Fig.~\ref{fig:dcrho}) and the temperature corresponding to a maximum resistivity.~\cite{palstra:86,mydosh:90,behnia:09} On the contrary, the optical resistivity is largely unaffected by the electron-electron contribution. Figure~\ref{fig:opt_ri_ee} shows the frequency dependence of the optical scattering rate at fixed temperature ($=22$ K) for the same values of $\alpha$ ($=0,1,2$) as in the {\em dc} case (Fig.~\ref{fig:dcrho}). As it is obvious from the figure, $1/\tau(\Omega,T)$ is practically the same for all three values of $\alpha$, except for a small overall shift. Repeating the same procedure as was applied to the numerical data in Figs.~\ref{fig:res_omega} and \ref{fig:intercept}, we again arrive at the result that coefficient $b$ in Eq.~(\ref{tau_exp}) is very close to $1$.

\begin{figure}[t]
\includegraphics[width=0.4
\textwidth]{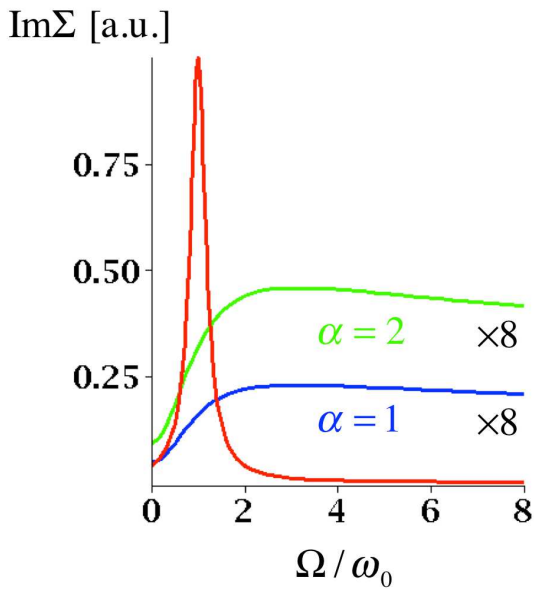}
\caption{(color on-line). Resonant-level (a sharply peaked curve) and electron-electron contributions to the imaginary part of the self-energy.
The electron-electron parts correspond to $\alpha=1,2$ and were multiplied by a factor of $8$ for clarity.}
\label{fig:sigma_ri_ee}
\end{figure}

The results presented above indicate that the electron-electron and resonant-level contributions to the self-energy affect
different parts of the frequency range: whereas the electron-electron contribution is largely responsible for the $T$ dependence of the {\em dc} resistivity and has practically no effect on the high-frequency optical resistivity, the resonant-level
 contribution
 determines the optical resistivity but plays only a secondary role in controlling the {\em dc} resistivity. This happens because {\em dc} and optical
measurements probe different parts of the electron spectrum (cf. Fig.~\ref{fig:sigma_ri_ee}).  At sufficiently low temperatures, i.e., at $T\ll \omega_0$, a {\em dc} measurement probes the spectrum in the region $\omega\sim T\ll\omega_0$, where both contributions to the self-energy
vary smoothly with $\omega$ (as $\omega^2$). The resulting $T$ dependence of $\rho(0,T)$ is just a sum of two $T^2$ terms with opposite signs, and the electron-electron contribution wins this competition rather easily. On the other hand,
the optical scattering rate is controlled by the region $\omega\sim\omega_0$, where the resonant-level contribution has a sharp peak and thus dominates over the electron-electron one, even if the latter is strong enough  to control the {\em dc} resistivity.

Concluding this section, we would like to emphasize the importance of a sharp feature in the elastic contribution to the self-energy, $\Sigma_{\mathrm{el}}(\omega)$. Indeed, the classification scheme of different behaviors of $1/\tau(\Omega,T$) based on the magnitude and sign of the coefficient $a$, as defined by Eq.~(\ref{taylor}), would predict substantially different values of $b$ for the values of the parameter $\alpha$ used above. It is easy to see that, in the resonant scattering model, $a=3/\alpha$ which, according to Eq.~(\ref{b_def}), implies that
$b=(4\alpha+3)/(\alpha+3)$. This  formula gives $b=1$ for $\alpha=0$; $b=7/4\approx 1.75$ for $\alpha=1$; and
$b=11/5\approx 2.2$ for $\alpha=2$. Nevertheless, fitting $1/\tau(\Omega,T)$ curves computed with a full form of $\Sigma_{\mathrm{ee}}$ rather than with its Taylor expansion, we obtained $b\approx 1$ in all of these cases. The reason for this discrepancy is that the Taylor expansion is not applicable near a sharp peak $\Sigma_{\mathrm{ee}}$, and one needs to use the classification scheme based on Eqs.~(\ref{tau}) and (\ref{b_def}) with certain care.

\section{summary}
\label{sec:7}
The main purpose of this paper was to highlight the universality of the FL result for the optical resistivity, Eq.~(\ref{gurzhi}).
We showed that, within the Kubo formalism which takes full account of vertex corrections to the conductivity, Eq.~(\ref{gurzhi}) holds for an arbitrary lattice
  and for any form of the electron-electron interaction,
  as long as the system remains a FL and is away from nesting and Van Hove singularities.
In fact, the optical resistivity turns out to be more universal than the single-particle self-energy: whereas the latter is described by the conventional form given by Eq.~(\ref{cfl})
 only in canonical FLs, i.e., in $D>2$, and deviates
from this form in non-canonical FLs
i.e.,
in
in $1<D\leq 2$, the former is given by Eq.~(\ref{gurzhi})
both for canonical and non-canonical FLs. We showed that a particular scaling form in Eq.~(\ref{gurzhi}) takes its roots in analytic properties of the optical conductivity along the Matsubara axis and is
 consistent
with
 the bosonic first-Matsubara-frequency rule.

If a system
 contains
 not only of itinerant electrons but
 also localized degrees of freedom (magnetic moments or resonant levels), the functional form of the optical resistivity changes, as specified by Eq.~(\ref{tau_exp}). The
 magnitude of the coefficient $b$ in this equation
 depends on the interplay between
 inelastic (electron-electron) and elastic
 scattering mechanisms.
Completely inelastic electron-electron
 scattering
corresponds to $b=4$;
completely elastic scattering from, e.g., resonant
levels, gives $b=1$; intermediate cases, where elastic and inelastic channels are mixed, correspond to $1<b<
\infty$.

As far as the existing experiments are concerned, the value of $b=4$ has never been reported. In  some cases, including the latest detailed study of the optical conductivity in \ho~(Ref.~\onlinecite{timusk11}), the coefficient $b$ has been found to be close to $1$, which indicates a completely elastic scattering mechanism;
a recent study of the Hg-based underdoped cuprate reports $b\approx 2.3$; yet another study of the BEDT-TTF organic material reports $b\approx 5.6$.
We considered a simple model of scattering from resonant levels, and showed
it
  is capable of reproducing the major features of the optical resistivity in \ho\;  above $T_c^{\mathrm{HO}}$. On the other hand, the $T$ dependence of the {\em dc} resistivity can only be explained in a model which combines elastic and inelastic electron-electron scattering mechanism. We deliberately refrained from identifying a microscopic nature of resonant levels, except for stating that they are not likely to be extrinsic resonant impurities. More likely, deep $f$ states of rare earth atoms play the role of incoherent resonant scatterers at rather high energy scales probed in optical measurements.

 If this picture is correct, it tells us something new about a crossover between coherent and incoherent regimes in heavy-fermion materials. The conventional
 scenario of this crossover
  is that the only energy scale is the Kondo temperature ($T_K$). Above $T_K$, localized magnetic moments scatter electrons incoherently, as in diluted Kondo alloy. Below $T_K$, a (heavy) FL state is formed and localized moments do not scatter electrons anymore but participate in formation of a coherent Bloch state. The low-energy FL state is supposed to have all the attributes of a standard FL, in particular, the coefficient $b$ must be equal to $4$.
  This
  scenario
   is probably correct as long as the evolution of the system is traced along the temperature axis. Optical measurements add one more dimension: frequency. In the presence of elastic scattering, the variations of temperature and frequency do not have the same physical consequences because the scattering cross-section depends on the electron energy, and thus on the frequency of light, but not on the temperature. It appears that the crossover between the incoherent and coherent regimes along the frequency axis contains an intermediate interval, where localized states scatter itinerant electrons neither as Kondo spins nor as screened Kondo clouds but rather as resonant levels.

Regardless of the validity
of
a particular
model for
elastic scattering,
 we hope that our paper will help to recognize the
 importance
of the
 interplay between $\Omega$ and $T$ dependencies
 in the optical data.
 We believe that, on par with much studied recently Wiedemann-Franz law which, if satisfied, indicates
not only the FL nature of the ground state but also
 complete elasticity of the
 underlying  scattering mechanism, systematic studies of the coefficient $b$ can tell us
 something new
  about the interplay between elastic and inelastic channels in strongly correlated electron systems.

\acknowledgments
We are particularly thankful to T. Timusk for motivating us to perform this study, and  to all authors of Ref.~\onlinecite{timusk11} for allowing us to use their data in our paper.
Helpful discussions with
D. Basov,
M. Broun,
D. Dessau,
P. Coleman,
 S. Dodge,
 M. Dressel,
 A. Georges,
K. Ingersent,
Y.-B. Kim,
P. Kumar,
 M. Kennett,
 D. van der Marel,
A. Millis,
  U. Nagel, T. Room, M. Sheffler,
 D. Tanner, A.-M. Tremblay, and V. I. Yudson are gratefully acknowledged. The work was supported by NSF-DMR 0906953 and Humboldt foundation (A. V. Ch.), and by NSF-DMR  0908029.
 We are thankful to
 MPIPKS Dresden (A.V. Ch. and D.L.M), the Aspen Center of Physics (A. V. Ch.), the Ruhr-University Bochum (A.V. Ch.),
 Simon Fraser University (D.L.M.), and Swiss NSF
\lq\lq QC2 Visitor Program\rq\rq\/ at
 the University of Basel (D.L.M.)
 for hospitality during the various phases of this work.
 The Aspen Center of Physics is supported in part by the NSF Grant 1066293.

\end{document}